\documentclass[%
 reprint,
superscriptaddress,
 amsmath,amssymb,
 aps,
 pra,
]{revtex4-2}
\usepackage{float}
\usepackage{graphicx}% Include figure files
\usepackage{dcolumn}% Align table columns on decimal point
\usepackage{bm}% bold math
\usepackage{hyperref}% add hypertext capabilities

\usepackage[T1]{fontenc}
\usepackage[utf8]{inputenc}
\usepackage{amsmath}
\usepackage{amssymb}
\usepackage{amsfonts}

\usepackage{gensymb}

\newcommand{\zh}{\bm}
\newcommand{\real}{\mathop{\rm Re}\nolimits}
\newcommand{\image}{\mathop{\rm Im}\nolimits}

\newcommand{\dee}{{\varepsilon}}

\newcommand{\zhr}{{\zh r}}

\newcommand{\zhe}{{\zh e}}
\newcommand{\zhn}{{\zh n}}
\newcommand{\zhp}{{\zh p}}
\newcommand{\zhq}{{\zh q}}
\newcommand{\zhk}{{\zh k}}
\newcommand{\zhnu}{{\zh\nu}}

\newcommand{\zhI}{{\zh I}}

\newcommand{\zhS}{{\zh S}}

\newcommand{\zhalpha}{{\zh\alpha}}

\newcommand{\zhnabla}{{\zh\nabla}}
\newcommand{\zhmu}{{\zh\mu}}

\newcommand{\zhsigma}{{\zh \sigma}}

\newcommand{\zhzeta}{{\zh \zeta}}

\newcommand{\hI}{{\hat{I}}}

\newcommand{\xl}{{l'}}

\newcommand{\xm}{{\overline{m^{\vphantom{0}}}}}

\newcommand{\Br}[1]{(\ref{#1})}
\newcommand{\Eq}[1]{Eq.\ (\ref{#1})}

\newcommand{\Eqss}[2]{Eqs.\ (\ref{#1},\ref{#2})}

\newcommand{\txt}[1]{{\rm #1}}

\newcommand{\mx}{{m_e}}
\newcommand{\trace}{\mathop{\rm Tr }\nolimits}
\newcommand{\threej}[6]{
        \left(
        \begin{array}{ccc}
        #1  & #2  & #3 \\
        #4  & #5  & #6 \\
        \end{array}
        \right)
        }

\newcommand{\vectorb}[2]{
\left(
\begin{array}{c}
#1\\
#2\\
\end{array}
\right)
        }

\begin{document}

%\preprint{}

%\title{Investigation of  nuclear structure corrections for electron and muon elastic scattering on bare nuclei}

\title{Hyperfine-driven polarization effects in elastic muon scattering by atomic nuclei}


%\author{one}
%\thanks{}
%\affiliation{}

\author{Ivan A. Moiseev}
%\email{o.y.andreev@spbu.ru}
\affiliation{St. Petersburg State University, 7/9 Universitetskaya nab., St. Petersburg, 199034, Russia}
\affiliation{Petersburg Nuclear Physics Institute named by B.P. Konstantinov of National Research Centre ''Kurchatov Institute'', Gatchina, 188300, Leningrad District, Russia}
\author{Daria M. Vasileva}
%\email{}
\affiliation{Petersburg Nuclear Physics Institute named by B.P. Konstantinov of National Research Centre ''Kurchatov Institute'', Gatchina, 188300, Leningrad District, Russia}
\author{Konstantin N. Lyashchenko}
%\email{}
\affiliation{St. Petersburg State University, 7/9 Universitetskaya nab., St. Petersburg, 199034, Russia}
\affiliation{Petersburg Nuclear Physics Institute named by B.P. Konstantinov of National Research Centre ''Kurchatov Institute'', Gatchina, 188300, Leningrad District, Russia}
\author{Igor A. Zhosan}
%\email{o.y.andreev@spbu.ru}
\affiliation{St. Petersburg State University, 7/9 Universitetskaya nab., St. Petersburg, 199034, Russia}
\affiliation{Petersburg Nuclear Physics Institute named by B.P. Konstantinov of National Research Centre ''Kurchatov Institute'', Gatchina, 188300, Leningrad District, Russia}
\affiliation{
St. Petersburg State Technological Institute (Technical University), 26 Moskovski ave., St. Petersburg, 190013, Russia
}
\author{Deyang Yu}
%\email{Corresponding author. Email: d.yu@impcas.ac.cn}
\affiliation{ Institute of Modern Physics, Chinese Academy of Sciences, Lanzhou, China}
\affiliation{ University of Chinese Academy of Sciences, Beijing, China}
\author{Oleg Yu. Andreev}
%\email{o.y.andreev@spbu.ru}
\affiliation{St. Petersburg State University, 7/9 Universitetskaya nab., St. Petersburg, 199034, Russia}
\affiliation{Petersburg Nuclear Physics Institute named by B.P. Konstantinov of National Research Centre ''Kurchatov Institute'', Gatchina, 188300, Leningrad District, Russia}


\date{\today}

\begin{abstract}
We present a theoretical study of polarization dynamics in the elastic scattering of muons and antimuons by high-magnetic-moment atomic nuclei, using ${}_{83}^{209}$Bi as a benchmark. The analysis tracks the evolution of (anti)muon polarization and its transfer to the nucleus, emphasizing nuclear-structure corrections. These corrections significantly alter both the differential cross sections and the polarization observables. Our results demonstrate that measuring scattered muon polarization serves as a sensitive probe for nuclear charge radii and magnetic dipole moments.
\end{abstract}
%Furthermore, the polarization transfer mechanism offers a route for obtaining polarized nucleus.

%\keywords{}

\maketitle


\section{Introduction}
The scattering of charged leptons (electrons or muons) by atomic nuclei constitutes one of the most fundamental processes in atomic, nuclear, and plasma physics, with wide-ranging applications spanning astrophysical environments and laboratory plasma diagnostics \cite{Walecka_2023}. In this work, we focus on the low-to-intermediate relativistic regime characterized by collision velocities $\beta \le 0.8$ (where $\beta = v/c$), using ${}_{83}^{209}$Bi as a benchmark. For an electron, this velocity domain corresponds to a kinetic energy of $\dee^{\txt{kin}(e)} \le 0.34\,\mbox{MeV}$, whereas for the significantly heavier muon, it corresponds to $\dee^{\txt{kin}(\mu)} \le 70\,\mbox{MeV}$. 

In this energy domain, the elastic scattering of electrons by point-like nuclei and the corresponding dynamics of electron polarization have been exhaustively documented. However, the associated nuclear structure corrections to the differential cross section have received far less attention. For electrons at these velocities, the de Broglie wavelength is relatively large compared to the nuclear dimensions, meaning the nucleus is perceived essentially as a point charge; thus, finite-size and structural corrections remain exceedingly small and are routinely neglected in standard calculations. Significant electron-nuclear structure corrections typically only manifest within specific resonance channels, such as those proceeding via the formation and subsequent autoionization of bound electronic states \cite{andreev08pr,andreev09p042514,Vasileva2021PhysRevA.104.052808,Vasileva_2024}.

For equivelocity muons and antimuons, the physical landscape is fundamentally altered. Muons are structurally identical to electrons but possess a mass roughly 207 times greater ($m_{\mu} \approx 207 m_{e}$). This large mass scales down the characteristic spatial extent of the lepton's wave function (the effective Bohr radius), forcing the muon to probe regions significantly closer to, and even within, the nuclear boundary \cite{Knyazeva2022PhysRevA.106.012809}. Consequently, the muon's interactions are extraordinarily sensitive to the internal distributions of nuclear charge and magnetism. 

Recent developments in accelerator physics and muon source technology \cite{GORRINGE201573,Cline2022PhysRevC.105.055201,Bao_2023,Cai2024PhysRevAccelBeams.27.023403} have yielded substantial improvements in beam quality, intensity, and polarization control. These experimental breakthroughs have sparked intensive investigations into bound muonic systems, revealing unique muonic effects in atomic physics and establishing highly precise pathways for extracting core nuclear parameters \cite{Adamczak2018refId0,Antognini2020PhysRevC.101.054313,Paul2021PhysRevLett.126.173001,Knyazeva2022PhysRevA.106.012809,andreev2025cpl,andreev2026muonenergy}.

Conventionally, nuclear ground-state properties, such as the root-mean-square charge radius and the magnetic dipole moment, are extracted either via high-energy electron scattering ($E^{(e)} > 50\,\mbox{MeV}$) or through high-precision laser and X-ray spectroscopy of electronic and muonic atoms \cite{Hofstadter1956RevModPhys.28.214,Barrett_1974,DEVRIES1987495,Skripnikov2018PhysRevLett.120.093001,stone_2025_16kpj-2k407,Gustafsson2025wbdx-k3cd}.
In this work, we demonstrate that direct scattering of free, polarized muons offers an alternative paradigm: by measuring joint lepton-nucleus polarization dynamics rather than conventional unpolarized cross sections, the nuclear magnetic moment can be extracted cleanly through its distinct footprint on the spin polarization observables.

The primary objective of this paper is to explore the distinct features of free muon-nucleus scattering and systematically compare the scattering behavior of free muons against their electronic counterparts. The foundational relativistic framework for lepton scattering by point-like targets, including analytical expansions in powers of the fine-structure constant scaled by the nuclear charge $(\alpha Z)$, was established in seminal works \cite{mott29,mott32,McKinley1948PhysRev.74.1759,Doggett1956PhysRev.103.1597,johnson1961}. Furthermore, the theoretical treatment of polarization transfer and polarization dynamics during the scattering process was outlined by Tolhoek \cite{Tolhoek1956}, with comprehensive numerical evaluations of the scattering asymmetry, referred to as the Sherman function, provided in \cite{sherman56}. 

Building upon these foundations, our study introduces a detailed analysis of non-point-like corrections. The main sources of deviations between muon and electron scattering mechanisms are finite nuclear size effects and magnetic hyperfine interactions between the continuum lepton and the nuclear magnetic moments. In the present work, we comprehensively investigate the scattering of both muons and antimuons by bare nuclei, paying particular attention to how their interactions with the nuclear magnetic dipole moment influence the scattering profiles and the joint polarization dynamics of the lepton-nucleus system. We discuss the explicit sensitivity of the spin polarization dynamics to the charge radius and the magnetic moment of the nucleus.

Relativistic units with the fine-structure constant $\alpha=e^2/(\hbar c)$ are used throughout the paper.

%%%%%%%%%%%%%%%%%%%%%%%%%%%%%%%%%%%%%%%%%%%%%%%%%%%%%%%%%%%%%%%%%%
\section{Theory}
We consider the elastic scattering of a lepton (electron, positron, muon, or antimuon) by a bare nucleus. Our theoretical treatment encompasses both the differential cross section and the associated polarization dynamics of the lepton-nucleus system. Particular emphasis is placed on nuclear-structure corrections, which we decompose into two main contributions: finite-size effects arising from the spatial distribution of nuclear charge, and electromagnetic hyperfine interactions originating from the nuclear magnetic dipole moment.

\subsection{Wave function}
The wave function of a lepton (electron, positron, muon, antimuon) satisfies the Dirac equation
\begin{eqnarray}
(\zhalpha\hat{\zhp}+m_{\ell}\beta + V)\psi
&=&\label{diraceq}
\dee\psi
\,,
\end{eqnarray}
where $\hat{\zhp}$ is the momentum operator, $\zhalpha$ and $\beta$ are the Dirac matrices, $m_{\ell}$ is the lepton mass.
The potential is given by
\begin{eqnarray}
V
&=&\label{diracv}
q\frac{\alpha Z}{r} + \Delta V
\,,
\end{eqnarray}
where $q$ is the charge sign of the lepton ($q=-1$ for electron and muon, and $q=+1$ for positron and antimuon).
The first term describes Coulomb interaction with point-like nucleus and the second term ($\Delta V$) takes into account a local scalar potential that can account for nuclear size corrections and the Uehling correction \cite{uehling35PhysRev.48.55,Fullerton1976PhysRevA.13.1283,mohr1998pr293-227}.

The scattering process is considered in the Furry picture \cite{furry51}, where the interaction of the lepton with the electric field of the atomic nucleus is treated exactly. The in- and out-wave functions, describing the incident and scattered lepton with energy $\dee$, momentum $\zhp$, and polarization $\mu$, can be expressed as \cite{akhiezer65b}
\begin{eqnarray}
\label{psior}
\psi^{(\pm)}_{\dee,\zhp\mu}(\zhr)
&=&\nonumber\frac{(2\pi)^{3/2}}{\sqrt{p\varepsilon}} 
\\
&&\times\sum_{jlm}\Omega^{+}_{jlm}(\zh\nu)\upsilon_{\mu}e^{\pm i \phi_{jl}}i^{l}\psi_{\varepsilon jlm}(\zhr)
\,,
\end{eqnarray}
where $p=|\zhp|$, $\zhnu=\zhp/p$, $\phi_{jl}$ are the phase shifts, and $\Omega_{jlm}$ is a spherical spinor \cite{Varshalovich1988QuantumTO}.
The wave functions $\psi_{\dee,jlm}$ satisfy \Eq{diraceq} and describe the lepton state with certain energy $\dee$, angular momentum $j$, its projection $m$ and parity $(-1)^l$.
The spinor $\upsilon_{\mu}=\upsilon_{\mu}(\zhnu)$ is defined by
\begin{eqnarray}
\frac{1}{2}( {\zh\nu\sigma})\upsilon_{\mu}(\zhnu)
&=&\label{spinor}
\mu\upsilon_{\mu}(\zhnu)
\,,
\end{eqnarray}
where $\zhsigma$ are the Pauli matrices.

For the case of point-like nucleus ($\Delta V=0$), the wave functions $\psi_{\dee,jlm}$ and phase shifts $\phi_{jl}$ can be obtained analytically \cite{akhiezer65b} (see also Appendix~\ref{appendix-fsm}). For $\Delta V\neq 0$, when nuclear size or Uehling corrections are included, the wave functions and phase shifts are computed numerically using the Salvat package \cite{salvat95}.

\subsection{Coulomb scattering amplitude}
We refer to the amplitude of scattering by the local scalar potential in Eq.~\eqref{diracv}, where $\Delta V$ is a short-range potential, as the Coulomb scattering amplitude. This amplitude can be written as
\begin{eqnarray}
U^{\textrm{Coul}}_{\mu_{i}\mu_{f}}(\theta,\varphi)
&=&\label{eqn260702n01}
\frac{-2\pi}{\varepsilon}\sum_{m}[\upsilon_{\mu_{f}}({\zhnu}_{f})]^{*}_{m}M^{\textrm{Coul}}_{m\mu_{i}}(\theta,\varphi),
\end{eqnarray}
where $\theta$ and $\phi$ are angles of scattered particle's momentum ($\zhnu_f(\theta,\varphi)$), with $z$-axis aligned along the initial momentum direction ($\zhp_i=\zhe_z p_i$),
$\mu_i$, $\mu_f$ are the initial and final polarization of the scattered lepton.
The Coulomb scattering matrix is given by
\begin{eqnarray}
M^{\textrm{Coul}}(\theta,\varphi)
&=&\label{Mmatrix}
\left(
\begin{array}{cc}
f(\theta)&g(\theta)e^{-i\varphi}\\
-g(\theta)e^{i\varphi}&f(\theta)\\
\end{array}
\right),
\end{eqnarray}
where the functions $f(\theta)$ and $g(\theta)$ can be defined in terms of phase shifts $\phi_{jl}$ \cite{landau4,burke2011b,res2020}
\begin{eqnarray}
 f(\theta)
&=&\label{f}
\frac{1}{2\pi i}\sum_{jl}|\varkappa|(e^{2i\phi_{\varkappa}}-1)P_{l}(\cos{\theta}),
\end{eqnarray}
\begin{eqnarray}
 g(\theta)
&=&\label{g}
\frac{1}{2\pi i}\sum_{l}(e^{2i\phi_{\varkappa=-l-1}}-e^{2i\phi_{\varkappa=l}})P^{1}_{l}(\cos{\theta})
\,.
\end{eqnarray}
Here $\varkappa=(j+\frac{1}{2})(-1)^{j+l+1/2}$ represents the Dirac angular quantum number, with $\phi_{\varkappa}\equiv \phi_{jl}$.
The function $P^{}_{l}(x)$ denotes the Legendre polynomial, while $P^{1}_l(x)=-\sqrt{1-x^2}\frac{d}{dx}P_l(x)$ is the associated Legendre polynomial.
The numerical computation of the $f(\theta)$ and $g(\theta)$ functions follows the approach described in \cite{res2020}.

\subsection{Nuclear size and vacuum polarization correction}
To account for the nuclear size correction (NS), we performed calculations using both a point-like nucleus potential (PN)
\begin{eqnarray}
 V^{\text{PN}}(r)
&=&
q\frac{\alpha Z}{r}
\end{eqnarray}
and a finite charge distribution (Fermi model \cite{ANGELI201369})
\begin{eqnarray}
V^\txt{F}(r)
&=&\nonumber
q4\pi\alpha
\left(
\frac{1}{r}\int\limits^{r}_{0}d{r'}\,r'^2 \rho({r'})
\right.
\\
&&
\left.
+\int\limits^{\infty}_{r}d{r'}\,\,{r'}\rho({r'})
\right)
\,,
\end{eqnarray}
where $\rho(r)$ is the nuclear charge density
\begin{eqnarray}
\rho(r)
&=&
\frac{N}{1+\exp\left(\frac{r-c}{a}\right)}
\,.
\end{eqnarray}
The normalization constant $N$ is fixed by
\begin{eqnarray}
4\pi\int\limits^{\infty}_{0}dr\,r^2\rho(r)
&=&
Z
\,.
\end{eqnarray}
The skin parameter $a$ is set to $a=0.535\,\mbox{fm}$
and the half-density radius $c$ is determined from the root-mean-square radius
\begin{eqnarray}
r^\txt{rms}
&=&
\left(\frac{4\pi}{Z}\int\limits^{\infty}_{0}dr\,r^4\rho(r)\right)^{1/2}
\,.
\end{eqnarray}

The difference between the Fermi and point-nucleus potentials,
\begin{eqnarray}
\Delta V^{\text{NS}}(r)
&=&
V^{\text{F}}(r)
-V^{\text{PN}}(r)
\,,
\end{eqnarray}
defines the interaction responsible for the nuclear size correction.

Since $\Delta V^{\text{NS}}$ is a short-range potential, it can in principle be treated either perturbatively or nonperturbatively. To first order in perturbation theory, the correction to the amplitude due to $\Delta V^{\text{NS}}$ is
\begin{eqnarray}
\label{pertub}
\Delta U^{\txt{NS}}_{\mu_{i}\mu_{f}}
&=&
\langle\psi^{(-)\txt{Coul}}_{\dee,\zhp_{f}\mu_{f}}|\Delta V|\psi^{(+)\txt{Coul}}_{\dee,\zhp_{i}\mu_{i}}\rangle
\,,
\end{eqnarray}
where $\psi^{(\pm)\txt{Coul}}$ are the in- and out-wave functions computed with the point-like nucleus potential. However, we find that the perturbative approach is adequate only for electrons; for muons, the correction is large and the perturbation theory fails. We therefore used the perturbative method only to test our numerical procedures (e.g., to check the Lippmann-Schwinger equation). In general, the nuclear size corrections were treated nonperturbatively.

The vacuum polarization correction due to the electron loop \cite{mohr1998pr293-227} was included nonperturbatively via the Uehling potential \cite{uehling35PhysRev.48.55}, i.e., by adding it to $V$ in Eqs.~\Br{diraceq} and~\Br{diracv}. Because the Uehling potential depends on the nuclear charge density distribution, it was computed with the Fermi nuclear model using the approximations presented in \cite{Fullerton1976PhysRevA.13.1283}. We found that, for a finite-size nucleus modeled by the Fermi distribution, the Uehling potential yields a relatively small correction to the scattering amplitude for both electrons and muons. For muons, however, a point-like nucleus model yields a significantly enhanced Uehling contribution that noticeably affects the scattering amplitude.
Accordingly, for muons, the Uehling potential must incorporate also the nuclear size corrections.
%This contribution is largely canceled when the corresponding nuclear finite-size correction is included.

%%%%%%%%%%%%%%%%%%%%%%%%%%%%%%%%%%%%%%%%%%%%%%%%%%%%%%%
\subsection{Nuclear magnetic dipole moment}
Another important correction is caused by the interaction of the scattering particle with the nuclear magnetic dipole moment (NMD). We consider the nucleus as a point-like magnetic dipole, not considering the Bohr-Weisskopf correction, which is determined by the distribution of the magnetic dipole moment over the nucleus.

Interaction of electron (or muon) with point-like magnetic dipole moment is given by the potential \cite{landau1975V2} (in the Gaussian convention)
\begin{eqnarray}
\Delta V^\txt{NMD}
&=&\label{VNMD}
-q
\frac{\hat{\zhmu}[\zhr\times\zhalpha]}{r^3}
\,,
\end{eqnarray}
where $q$ is the charge sign of the scattering particle (see \Eq{diracv}),
$\hat{\zhmu}$ is operator of the nucleus magnetic dipole moment.
The latter can be presented as
\begin{eqnarray}
\hat{\zhmu}
&=&
\frac{\mu}{I}\mu_N \hat{\zhI}
\,,
\end{eqnarray}
where $\mu_N=|e|/(2m_p)$ is the nuclear magneton, $m_p$ is proton mass, $\hat{\zhI}$ is the nucleus spin operator and $I$ is the nucleus spin, $\mu$ is the nucleus magnetic dipole moment measured in nuclear magnetons.

The interaction of the scattering particle with the nucleus magnetic dipole moment is taken into account in the first order of the perturbation theory.
The corresponding correction to the scattering amplitude reads as
\begin{eqnarray}
\Delta U^\txt{NMD}_{M_{i}\mu_{i},M_{f}\mu_{f}}
\!\!\!\!&=&\!\!\label{mU}
\langle
\Psi^{}_{IM_f,\zhp_f\mu_f}|
\Delta V^\txt{NMD}
|\Psi^{}_{IM_i,\zhp_i\mu_i}\rangle
\,,
\end{eqnarray}
where lepton-nucleus system is described by
\begin{eqnarray}
|\Psi^{}_{IM,\dee\zhp\mu}\rangle
&=&
|\psi^{(\txt{nuc})}_{IM}\rangle\otimes
|\psi^{(\ell)}_{\dee,\zhp\mu}\rangle
\,.
\end{eqnarray}
Here, $|\psi^{(\ell)}_{\dee,\zhp\mu}\rangle$ is the ket-vector describing the lepton (given by \Eq{psior}) and $|\psi^{(\txt{nuc})}_{IM}\rangle$ is the ket-vector describing the nucleus with spin $I$ and its projection $M$.
We assume that
\begin{eqnarray}
\langle \psi^{(\txt{nuc})}_{IM}|\hat{\zhI}^2|\psi^{(\txt{nuc})}_{IM'}\rangle
&=&
I(I+1)\delta_{MM'}
\,,
\\
\langle \psi^{(\txt{nuc})}_{IM}|\hat{I}_z|\psi^{(\txt{nuc})}_{IM'}\rangle
&=&
M\delta_{MM'}
\,.
\end{eqnarray}
The correction to the scattering amplitude \Eq{mU} can be written as
\begin{eqnarray}
\Delta U^\txt{NMD}_{M_{i}\mu_{i},M_{f}\mu_{f}}
&=&\nonumber
-\frac{q\mu\mu_N}{I}
\sum\limits_{\eta} (-1)^{\eta}
\langle \psi^{(\txt{nuc})}_{IM_f}|\hat{I}_{\bar{\eta}}|\psi^{(\txt{nuc})}_{IM_i}\rangle
\\
&&\label{HFSnoIn02}
\times
\langle
\psi^{(-)(\ell)}_{\dee,\zhp_f\mu_f}|
\frac{[\zhr\times\zhalpha]_{\eta}}{r^3}
|\psi^{(+)(\ell)}_{\dee,\zhp_i\mu_{i}}\rangle
\,,
\end{eqnarray}
where the spherical coordinates ($\eta=0,\pm1$, $\bar\eta\equiv-\eta$) were used for representation of the scalar product
%(Varshalovich, p.12, \S 1.2.1., Eq.~(11) and p.42, \S 2.3.3. Eq.~(7)
\cite{Varshalovich1988QuantumTO}.
The calculation of the last matrix element in \Eq{HFSnoIn02} is discussed in Appendix \ref{appendix-fsm}.
The integration over the angular variables and summation over the spin variables as well as the evaluation of the matrix elements of the nucleus spin operator are presented in Appendix~\ref{appendix-angle}.

%%%%%%%%%%%%%%%%%%%%%%%%%%%%%%%%%%%%%%%%%%%%%%%%%%%%%%%%%%%%%%%%%
\subsection{Scattering amplitude}
The scattering amplitude of a lepton by an atomic nucleus, including the corrections discussed above, can be written as
\begin{eqnarray}
U_{M_{i}\mu_{i},M_{f}\mu_{f}}
&=&\nonumber
\left(U^{\txt{PN}}_{\mu_{i},\mu_{f}}+\Delta U^{\txt{NS}}_{\mu_{i},\mu_{f}}\right)
\delta_{M_{i},M_{f}}
\\
&&\label{ufull}
+ \Delta U^{\txt{NMD}}_{M_{i}\mu_{i},M_{f}\mu_{f}}
\,,
\end{eqnarray}
where $U^{\txt{PN}}$ is the scattering amplitude by the point-like nucleus potential [\Eq{diracv} with $\Delta V=0$], $\Delta U^{\text{NS}}$ is the correction due to the nuclear size (NS) and vacuum polarization (electron loop in the Uehling approximation), and $\Delta U^{\txt{NMD}}$ is the correction due to the NMD interaction between the lepton and the nucleus. Since the Uehling correction computed with the Fermi nuclear model gives a rather small contribution, it is not studied separately but is included together with the NS corrections.

The differential cross section can be written as
\begin{eqnarray}
\frac{d\sigma}{d\Omega_f}
&=&\label{dcsu}
\frac{\dee^2}{2(2I+1)(2\pi)^2}
\sum\limits_{{M_{i}M_{f}}\atop{\mu_{i}\mu_{f}}}
\left|U_{M_{i}\mu_{i},M_{f}\mu_{f}}\right|^2
\,,
\end{eqnarray}
where $\Omega_f$ is the solid angle for the momentum of the scattered lepton. We average over the initial projections and sum over the final projections.

To study the role of the NS and NMD corrections, we perform separate calculations. The scattering amplitude including the NS corrections (and the Uehling correction) is referred to as
\begin{eqnarray}
U^{\text{NS}}_{M_{i}\mu_{i},M_{f}\mu_{f}}
&=&
U^{\text{PN}}_{\mu_{i},\mu_{f}}\delta_{M_{i},M_{f}}
+
\Delta U^{\text{NS}}_{\mu_{i},\mu_{f}}\delta_{M_{i},M_{f}}
\,.
\label{UNS}
\end{eqnarray}
The amplitude obtained in the full calculation, including both NS and NMD corrections, is denoted by
\begin{eqnarray}
U^{\text{NS+NMD}}_{M_{i}\mu_{i},M_{f}\mu_{f}}
&=&
U^{\text{NS}}_{\mu_{i},\mu_{f}}\delta_{M_{i},M_{f}}
+
\Delta U^{\text{NMD}}_{M_{i}\mu_{i},M_{f}\mu_{f}}
\,.
\label{UNSNMD}
\end{eqnarray}
Finally, we also present results where the NMD correction is calculated using the analytical Coulomb wave functions $\psi^{\text{C}}_{\varepsilon,jlm}$ for a point-like nucleus [Eq.~\eqref{psiorc}]. This amplitude is denoted by
\begin{eqnarray}
U^{\text{NS+NMD(PN)}}_{M_{i}\mu_{i},M_{f}\mu_{f}}
&=&
U^{\text{NS}}_{\mu_{i},\mu_{f}}\delta_{M_{i},M_{f}}
+
\Delta U^{\text{NMD(PN)}}_{M_{i}\mu_{i},M_{f}\mu_{f}}
\,.
\label{UNSNMDPN}
\end{eqnarray}

%%%%%%%%%%%%%%%%%%%%%%%%%%%%%%%%%%%%%%%%%%%%%%%%%%%%%%%%%%%%%%%%%%%%%
\subsection{Polarization theory}
\label{section-pol}
One of key contributions to polarization changes in muon scattering is the magnetic dipole interaction with the nucleus — particularly relevant for high-spin, high-magnetic-moment nuclei. To account for this interaction, we now discuss scattering of spin-$1/2$ particles on a target of arbitrary spin $S$.

The polarization of the beam of particles with spin $1/2$ in its rest frame is described by the density matrix
\begin{eqnarray}
	\hat{\rho}^{(\ell)}
        &=&\label{eqn260710n01}
        \dfrac{1}{2}\left(1+\zhzeta\hat{\zhsigma}\right)
        \,,
\end{eqnarray}
where $\zhzeta$ is the polarization vector of the initial beam, $P=|\zhzeta|$ is the degree of polarization of the beam, $\hat{\zhsigma}$ is the vector consisting of Pauli matrices.

While in electron scattering the initial electron beams used in experiment are usually unpolarized ($\zhzeta^{(e)}_i=0$), the muon and antimuon beams obtained in experiment typically have polarization degree $P_i^{(\mu^\mp)}$ close to $1$. Muon beams are polarized along the beam momentum ($\zhzeta^{(\mu^{-})}_i=\zhp_i/|\zhp_i|$), and antimuon beams opposite the beam momentum ($\zhzeta^{(\mu^{+})}_i=-\zhp_i/|\zhp_i|$) \cite{Blundell2021muon}.

In this work, we assume that the nucleus is initially in a completely unpolarized state and the initial nucleus spin state is given by
\begin{eqnarray}
	\hat{\rho}^{(\txt{nuc})}_i=\dfrac{1}{2S+1}\hat{I}_{2S+1}\,,
\end{eqnarray}
where $S$ is the spin of the nucleus, $\hat{I}_{2S+1}$ is the $(2S+1)\times(2S+1)$ identity matrix.

We assume that in the initial state the lepton beam and the nucleus do not interact, therefore the density matrix of the initial state is
\begin{eqnarray}
	\hat{\rho_i}=\hat{\rho}^{(\txt{nuc})}_i\otimes\hat{\rho}^{(\ell)}_i\,. \label{ispindm}
\end{eqnarray}

The polarization changes in the scattering are completely described by the scattering matrix $\hat{M}$ with matrix elements \Eq{ufull}
\begin{eqnarray}
	{M}_{M_{i}\mu_{i},M_{f}\mu_{f}}(\theta, \varphi)
	&=&\label{eqn260818n01}
	U_{M_{i}\mu_{i},M_{f}\mu_{f}}(\theta, \varphi)
\,.
\end{eqnarray}

The density matrix of the final state of the system after scattering is given by
\begin{eqnarray}
	\hat{\rho}=\dfrac{1}{\trace\left(\hat{M}\hat{\rho_i}\hat{M}^\dagger\right)}\hat{M}\hat{\rho_i}\hat{M}^\dagger\,.\label{fspindm}
\end{eqnarray}
The normalization factor $\trace\left(\hat{M}\hat{\rho_i}\hat{M}^\dagger\right)^{-1}$ in \Eq{fspindm} is needed so that $\trace \hat{\rho}$ would remain equal to $1$.
The differential cross section can be written as (see \Eq{dcsu})
\begin{eqnarray}
\frac{d\sigma}{d\Omega_f}
&=&
\frac{\dee^2}{(2\pi)^2}
\trace\left(\hat{M}\hat{\rho_i}\hat{M}^\dagger\right)
\,.
\end{eqnarray}

Generally, the system of nucleus and lepton after scattering is spin correlated, therefore the final density matrix \Eq{fspindm} cannot be written in the form similar to \Eq{ispindm}. However, in this study we do not consider this correlation and assume that either scattered lepton beam or nucleus polarization is detected but not both. Therefore, we can describe the polarization change of the lepton beam and the nucleus separately.

In this case, the polarization of the lepton beam is described completely by the polarization vector. The polarization vector after scattering is given by
\begin{eqnarray}
		\zhzeta^{(\ell)}=\dfrac{\trace \left(\hat{M}\hat{\rho_i}\hat{M}^\dagger \hat{I}_{2S+1}\otimes\hat{\zhsigma}\right)}{\trace\left(\hat{M}\hat{\rho_i}\hat{M}^\dagger\right)}\,. \label{polvec}
\end{eqnarray}

The description of the lepton polarization change in scattering by a target with spin $1/2$ was studied in \cite{Burke1974,Vasileva2021PhysRevA.104.052808}.
In particular, it was shown that for any initial polarization of the lepton beam, the lepton polarization change is described by five parameters.

The present work extends this theory to arbitrary‑spin targets. In this case, two more parameters (for a total of seven) are needed to describe the polarization change. Notably, the parameter describing polarization change in an initially unpolarized system (the Sherman function) and the parameter describing the cross‑section asymmetry with respect to the azimuthal angle $\varphi$ no longer coincide.  We discuss the description of scattering by systems with arbitrary spin and the general formula for the polarization vector in detail in Appendix \ref{appendixpol}.

For the purpose of this work we assume that the initial state density matrix $\hat{\rho}_i$ is as discussed previously -- i.e., it corresponds to non‑interacting unpolarized nucleus and longitudinally polarized muon ($\mu^-$) or antimuon ($\mu^+$) beam. Therefore, we do not analyze all seven parameters. Instead, we present three projections $P_\zhe^{(\mu^\mp)}$ of the polarization vector after scattering onto the axes most relevant for the experiment: the direction of the incident beam momentum $\zhp_i$, the direction of the scattered beam momentum $\zhp_f$ and the direction perpendicular to the scattering plane $\zhe_\perp=\dfrac{\zhp_i \times \zhp_f}{|\zhp_i \times \zhp_f|}$.

In order to determine these projections, we need to calculate traces of two types.  These traces can be obtained from the matrix elements corresponding to the scattering amplitudes \Eq{eqn260818n01}, whose calculation was described in previous sections:
\begin{eqnarray}
	\trace\left(\hat{M}\hat{\rho}_i^{(\mu^\mp)}\hat{M}^\dagger\right)
	&=&\nonumber
\frac{1}{2S+1}
\\
&&
\times
\sum_{ {M_i M_f}\atop{ \mu_f} } |M_{M_i 1/2, M_f \mu_f}|^2 \label{tracet1}
\,,
\\
\trace\left(\hat{M}\hat{\rho}_i^{(\mu^\mp)}\hat{M}^\dagger\hat{\zhsigma}_k\right)
&=&\nonumber
\dfrac{1}{2S+1}
\sum_{{M_i M_f}\atop{\mu_f \mu'_f}}
(\zhsigma_k)_{\mu'_f \mu_f}
\\
&&\label{tracet2}
\hspace{-32pt}
\times
M_{M_i \pm 1/2, M_f \mu_f}
M^*_{M_i \pm 1/2, M_f \mu'_f}
\,.
\end{eqnarray}
We also use the fact that the absolute value of the amplitude $M_{M_i,\mu_i,M_f\mu_f}$  is invariant under simultaneous sign reversal of all spin projections $M_{i},\,M_{f},\,\mu_{i},\,\mu_{f}$. The amplitudes used in the above formulae must be calculated using the same spinor $v_\mu(\zhe_0)$ in \Eq{psior} for both incident and scattered muon wavefunctions:
\begin{eqnarray}
	\frac{1}{2}( {\zhe_0\sigma})\upsilon_{\mu}(\zhe_0)
	&=&
	\mu\upsilon_{\mu}(\zhe_0)
	\,.
\end{eqnarray}

The polarization state of the nucleus with spin $S$ is generally described by the $(2S+1) \times (2S+1)$ density matrix $\hat{\rho}^{(\txt{nuc})}$ defined by $4S(S+1)$ real parameters. This density matrix can be written as an expansion over polarization operators $\hat{T}_{LM}(S)$ with complex coefficients $r_{LM}=\trace(\hat{T}^\dagger_{LM}(S) \hat{\rho}^{(\txt{nuc})})$ \cite{Varshalovich1988QuantumTO}:
\begin{eqnarray}
	\hat{\rho}^{(\txt{nuc})}&=&\sum_{L=0}^{2S}\sum_{M=-L}^{L} r_{LM} \hat{T}_{LM}(S)\,. \label{polopexp}
\end{eqnarray}
Since our aim is to estimate the polarization gained by an initially unpolarized nucleus, we consider the average nuclear spin projection on the arbitrary axis $\zhe_0$:
\begin{eqnarray}
	\bar{S}_{\zhe_0}&=&\trace \left(\hat{\rho}(\hat{\zhS} \zhe_0 \otimes \hat{I}_2)\right)\,.
\end{eqnarray}
It is convenient to rewrite $\hat{T}_{00}(S)$ and $\hat{T}_{1M}(S)$ terms in \Eq{polopexp} as
\begin{eqnarray}
	\hat{\rho}^{(\txt{nuc})}
        &=&
        \dfrac{1}{2S+1}\left(\hat{I}_{2S+1}+\zh\xi^{(\txt{nuc})}\hat{\zhS}+\hat{T}_{L\geqslant 2}\right)\,,
\end{eqnarray}
where $\hat{T}_{L\geqslant 2}$ contains only terms with polarization operators $\hat{T}_{LM}(S)$ with $L\geqslant 2$. When determining the average spin projection on any given axis $\zhe_0$ we must calculate traces such as $\trace(\hat{T}_{LM}(S)\zhS)$. Considering that
\begin{eqnarray}
	\trace(\hat{T}_{LM}(S)\hat{T}_{L'M'}(S))=(-1)^M\delta_{LL'}\delta_{M-M'}
\end{eqnarray}
term $\hat{T}_{L \geqslant 2}$ does not yield a contribution and we obtain:
\begin{eqnarray}
	\trace(\hat{\rho}^{(\txt{nuc})}(\hat{\zhS}\zhe_0))
        &=&
        \dfrac{S(S+1)}{3}(\zhe_0 \zh\xi^{(\txt{nuc})})
        \\
        &=&
        S(\zhe_0 \zhzeta^{(\txt{nuc})}_{\overline{S}})\,,
\end{eqnarray}
where
\begin{eqnarray}
\zhzeta^{\txt{nuc}}_{\overline{S}}
&=&\label{eqn260710n02}
\dfrac{S+1}{3}\zh\xi^{(\txt{nuc})}
\end{eqnarray}
is a vector consisting of average spin projection on the chosen basis set of vectors normalized by the maximum possible spin projection $S$.
It is refered to as the normalized average nuclear spin.
Its absolute value $|\zhzeta^{\txt{nuc}}_{\overline{S}}|\leqslant 1$ is analogous to polarization degree of spin-1/2 case.

The trace required for the numerical calculation of average spin projections can be obtained using the following formula, with the matrix elements $M$ the same as in \Eqss{tracet1}{tracet2}:
\begin{eqnarray}
	\trace\left(\hat{M}\hat{\rho}_i^{(\mu^\mp)}\hat{M}^\dagger\hat{S}_k\right)
	&=&\dfrac{1}{2S+1}\sum_{\substack{M_i M_f \\ M'_f \mu_f}} M_{M_i \pm 1/2, M_f \mu_f} \nonumber \\
	&\times& M^*_{M_i \pm 1/2, M'_f \mu_f}(\hat{S}_k)_{M'_f M_f}\,.
\end{eqnarray}

The matrix elements of spin operator for arbitrary spin $S$ are given by \cite{Varshalovich1988QuantumTO}:
\begin{eqnarray}
	(\hat{S}_x)_{\mu \mu'}
&=&
\sqrt{\dfrac{S(S+1)}{2}}
\left[C^{S\mu}_{S\mu',1-1}-C^{S\mu}_{S\mu',11}\right]
\\
	(\hat{S}_y)_{\mu \mu'}
&=&
i\sqrt{\dfrac{S(S+1)}{2}}
\left[C^{S\mu}_{S\mu',1-1}+C^{S\mu}_{S\mu',11}\right]
\\
	(\hat{S}_z)_{\mu \mu'}
&=&
\sqrt{S(S+1)}\,
C^{S\mu}_{S1\mu',1 0}
\,,
\end{eqnarray}
where $C^{J_{12}M_{12}}_{j_1m_1,j_2m_2}\equiv\langle j_1m_1j_2m_2|J_{12}M_{12}\rangle$ are Clebsch-Gordan coefficients \cite{Varshalovich1988QuantumTO}.

%%%%%%%%%%%%%%%%%%%%%%%%%%%%%%%%%%%%%%%%%%%%%%%%%%
\section{Results and discussion}
We investigate the role of nuclear-structure corrections in the elastic scattering of leptons by a bare nucleus. For this purpose, we select the bismuth nucleus $^{209}_{83}\mathrm{Bi}$ as a representative heavy, stable nucleus with a large spin and a sizable magnetic dipole moment. This nucleus is described by a charge root-mean-square radius $r^{\txt{rms}} = 5.5211\,\mbox{fm}$ \cite{ANGELI201369}, nuclear spin $I = 9/2$, and magnetic moment $\mu = 4.092\,\mu_N$ \cite{Skripnikov2018PhysRevLett.120.093001,stone_2025_16kpj-2k407}.

\begin{figure}[h!]
\includegraphics[width=0.5\textwidth]   {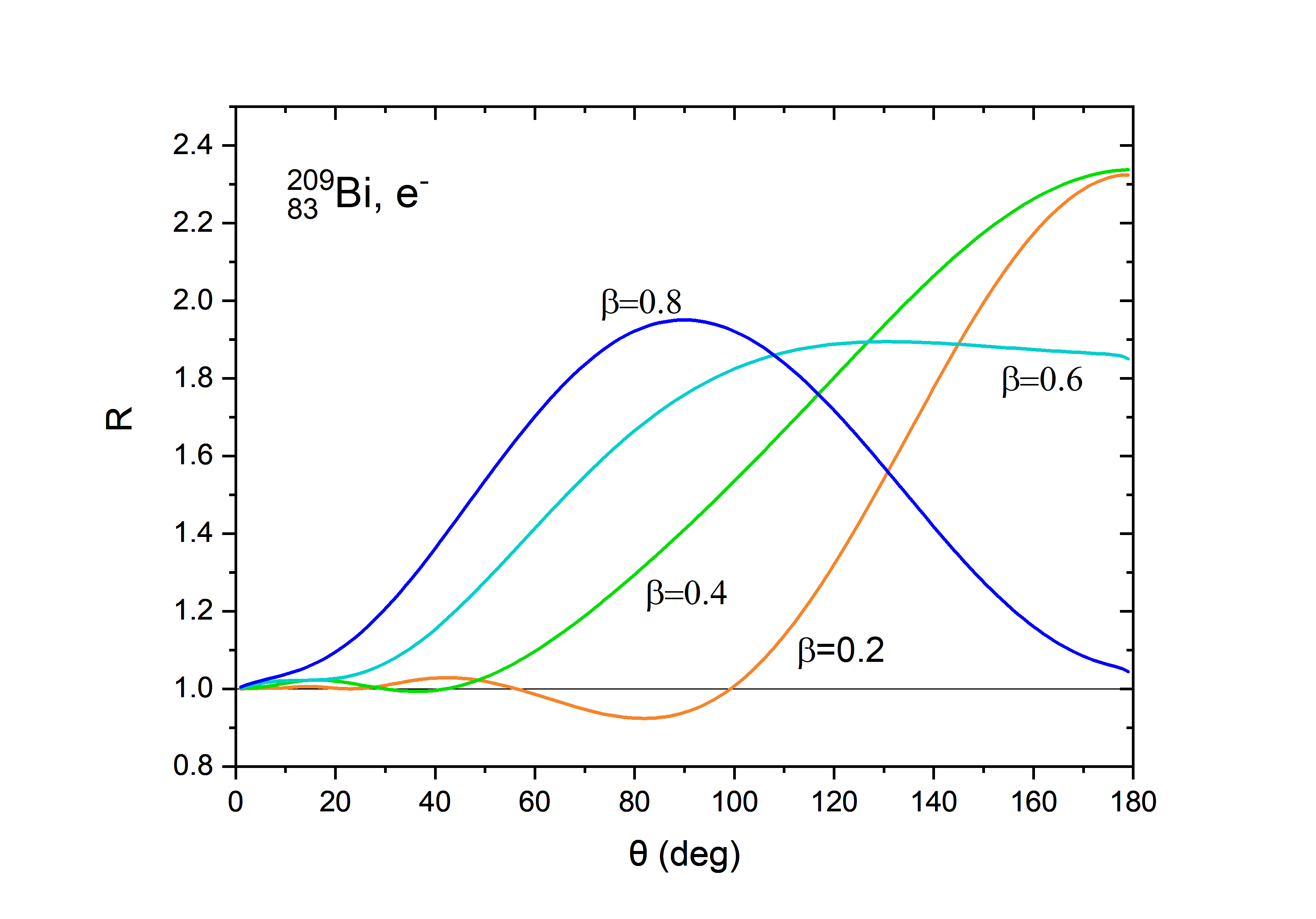}\\
\includegraphics[width=0.5\textwidth]   {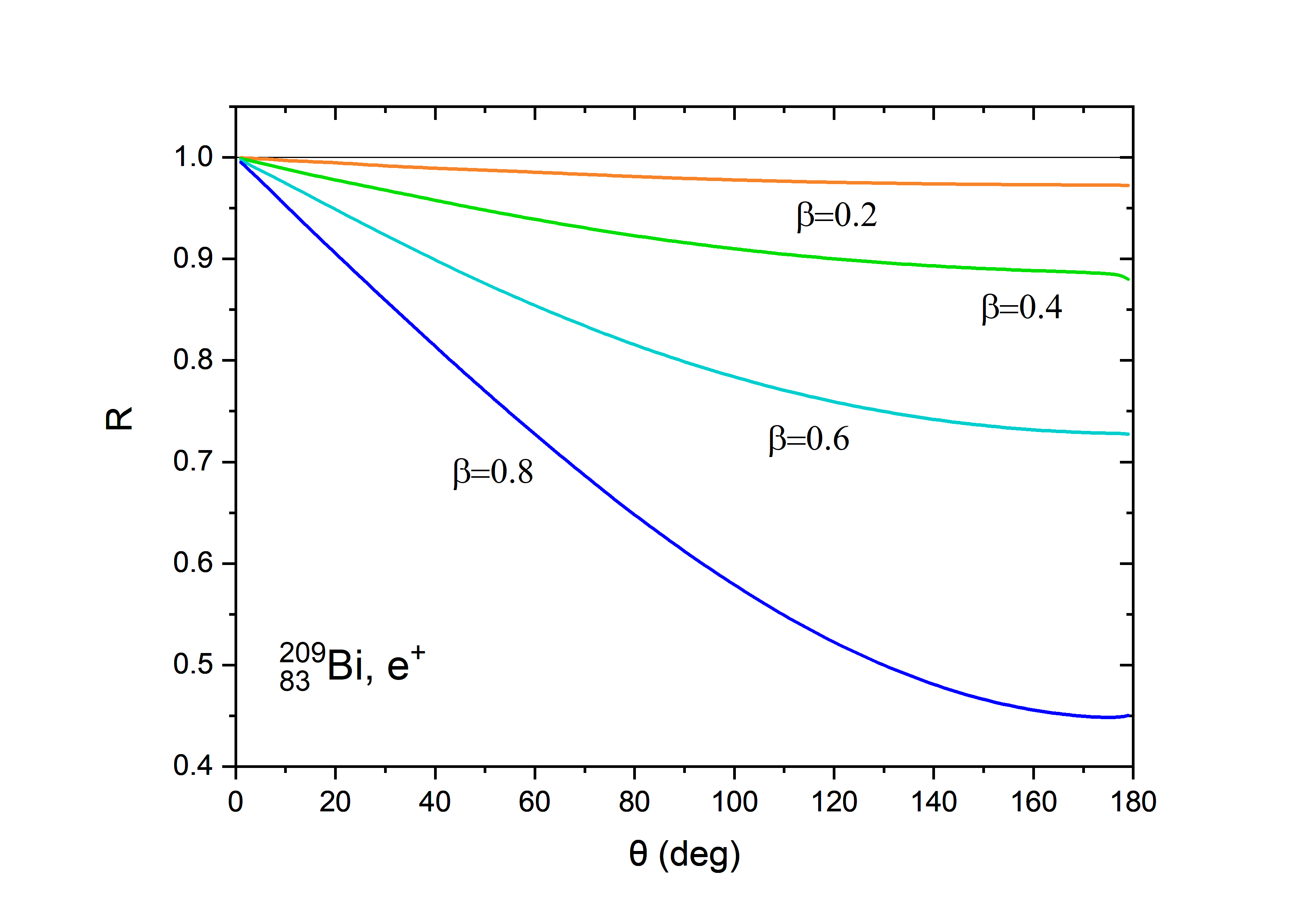}\\
\caption{The ratio $R(\theta)$ (see \Eq{ratio}) between the differential cross section and the Rutherford formula for the scattering of an electron (upper panel) and a positron (lower panel) by bare nuclei. The collision velocity $\beta$ is expressed in relativistic units.
The same ratios $R(\theta)$ apply to muons and antimuons when nuclear structure corrections are disregarded.}
\label{cs-electron}
\end{figure}

\begin{figure}[h!]
\includegraphics[width=0.5\textwidth]   {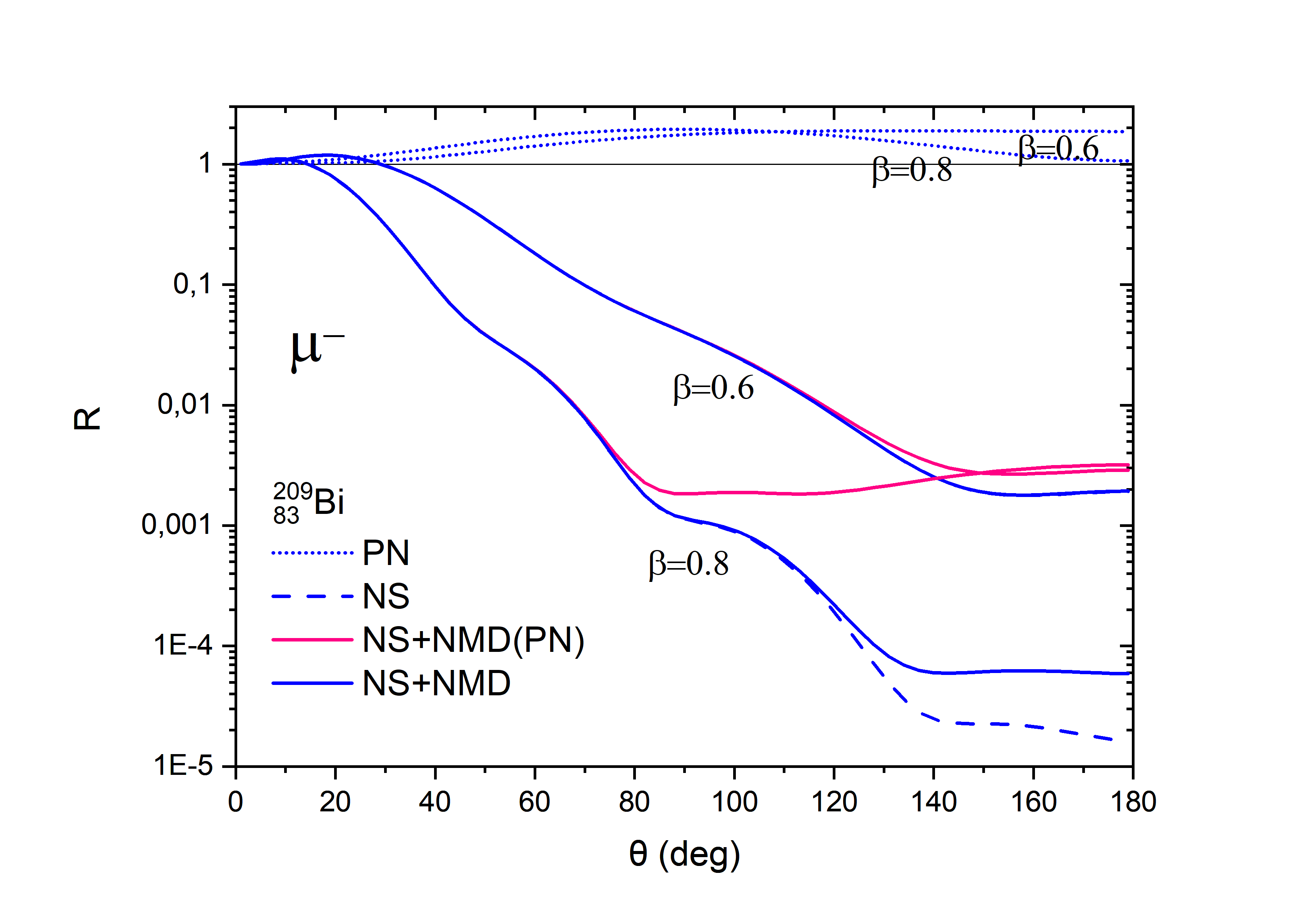}\\
\includegraphics[width=0.5\textwidth]   {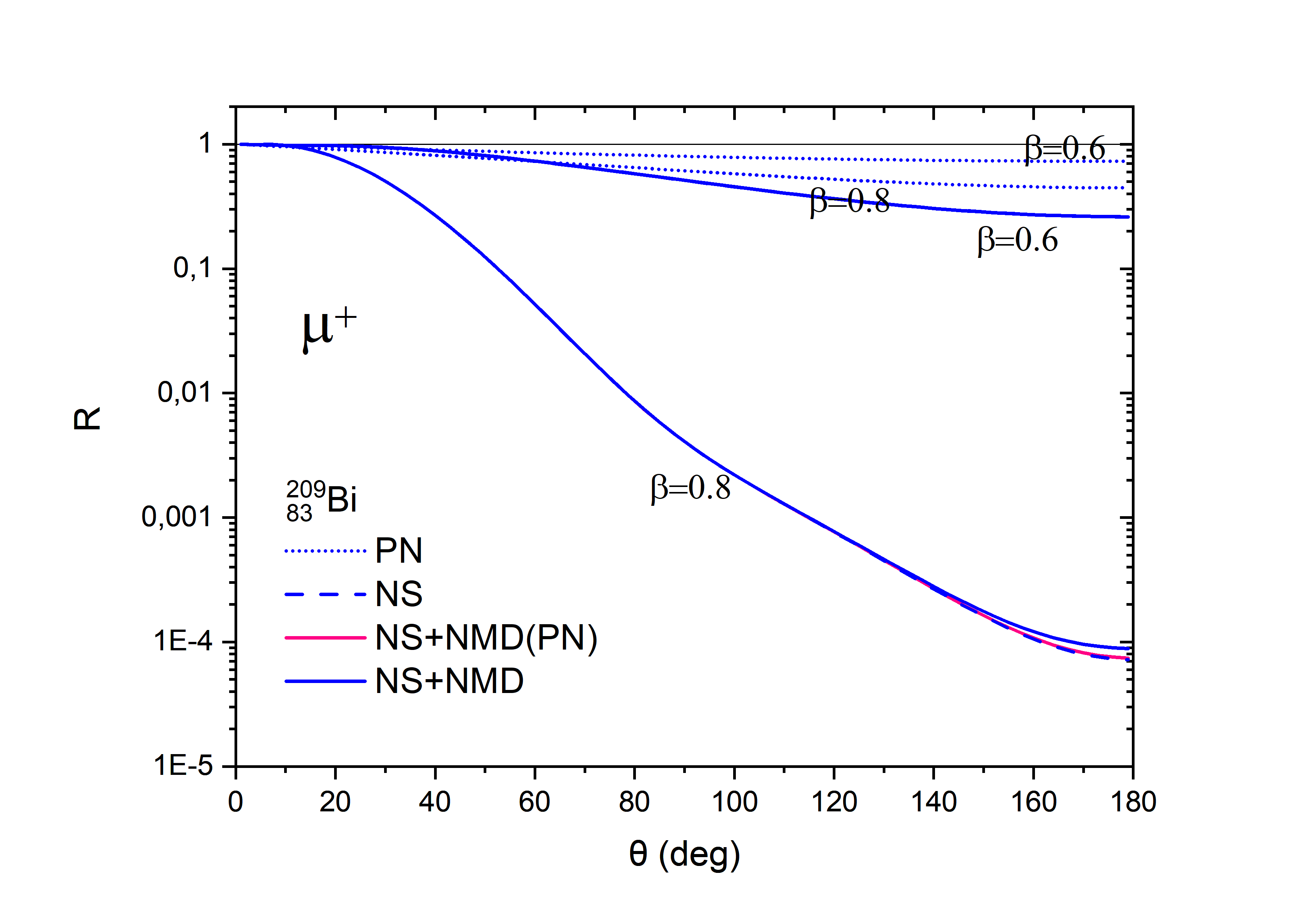}\\
\caption{The ratio $R(\theta)$ (see \Eq{ratio}) of the differential cross section to the Rutherford formula for scattering of a muon (upper panel) and an antimuon (lower panel) by a bare bismuth nucleus, for collision velocities $\beta=v/c$. The blue solid curves show the results of the full calculation (NS+NMD) including both the nuclear size and nuclear magnetic dipole corrections. The blue dotted curves correspond to scattering by a point-like nucleus, i.e., disregarding all nuclear structure corrections. The blue dashed curves include only the NS correction, omitting the NMD contribution. The red solid curves present the NS+NMD(PN) results, where the NMD correction is calculated using wave functions obtained for a point-like nucleus potential, while the other NS corrections are included.}
\label{cs-muon}
\end{figure}

%
% E= 10.53 keV = 19.1 MeV/u --> \beta=0.2
% E= 46.55 keV = 84.9 MeV/u --> \beta=0.4
% E= 127.75 keV = 233 MeV/u--> \beta=0.6
% E= 340.7 keV = 622 MeV/u --> \beta=0.8
%

\subsection{Differential cross section}
Nuclear structure corrections are typically negligible for electron scattering when the kinetic energy does not exceed several MeV. Previous studies have therefore focused primarily on relativistic effects \cite{Doggett1956PhysRev.103.1597,johnson1961}. For completeness, and to provide a baseline for comparison with our muon results, we present the full (averaged over initial projections and summed over final projections) DCS for electrons and positrons, normalized as the ratio $R$ to the nonrelativistic Rutherford formula
\begin{eqnarray}
R(\theta)
&=&\label{ratio}
\left.\frac{d^2\sigma(\theta)}{d^2\Omega}
\right/
\frac{d^2\sigma_{\txt{R}}(\theta)}{d^2\Omega}
\,,
\end{eqnarray}
where
\begin{eqnarray}
\frac{d^2\sigma_{\txt{R}}(\theta)}{d^2\Omega}
&=&
\left(\frac{\dee\alpha Z}{2p^2}\right)^2
\frac{1}{\sin^4(\theta/2)}
\,.
\end{eqnarray}

Figure~\ref{cs-electron} shows the results for electrons and positrons at collision velocities (in relativistic units) $\beta = 0.2$, $0.4$, $0.6$, and $0.8$, corresponding to the nuclear kinetic energies of $E = 19.2$, $84.8$, $233$, and $621$~MeV/u in the lepton rest frame, respectively. Deviations of the ratio  $R(\theta)$ from unity arise almost solely from relativistic corrections. Our results agree well with Ref.~\cite{Doggett1956PhysRev.103.1597}. Although the Rutherford formula is identical for electrons and positrons, the relativistic corrections make the DCSs for particle and antiparticle substantially different.
We note that within the PN approximation, when NS and NMD corrections are disregarded, the ratios  $R(\theta)$ for muons and antimuons are the same as those for electrons and positrons (presented in Fig.~\ref{cs-electron}), respectively.

In contrast, because the muon is about $200$ times heavier than the electron, nuclear structure corrections become significant at much lower velocities. Still, at small collision velocities ($\beta<0.6$), the nuclear structure corrections are relatively small, and the behavior of muons (antimuons) resembles that of electrons and positrons. Therefore, we limit our consideration to $\beta= 0.6$ and $0.8$.

Figure~\ref{cs-muon} presents the DCS for muon and antimuon scattering by a bismuth nucleus. The blue solid curves show the results of the full calculation, where both NS and NMD corrections are taken into account. The blue dotted curves give the results where nuclear structure corrections are neglected; they thus represent scattering by a pure Coulomb potential. These dotted curves coincide with the ratios $R(\theta)$ for electrons and positrons presented in Fig.~\ref{cs-electron}.

The blue dashed curves show the results where only the NS corrections are taken into account. This correction arises from the fact that the muon moves in a smeared nuclear charge distribution, which diminishes the electric field of the nucleus compared to the point-like Coulomb potential. As a result, the dashed curves are considerably lower (by several orders of magnitude) than those for the point-like Coulomb potential. The difference between the blue solid curves and the blue dashed curves gives the contribution of the NMD correction. Notably, the interaction of the muon with the nuclear magnetic dipole moment becomes significant at large scattering angles and can enhance the DCS severalfold.

The red solid curves show the results where the matrix elements corresponding to the NMD corrections were calculated using electron wave functions obtained for the pure Coulomb potential. They thus represent the importance of NS corrections for the calculation of the NMD corrections. When the NMD contribution is computed assuming a point-like nucleus (red solid curve), it overestimates the true NMD effect by a factor of several compared to the full NS+NMD calculation. This demonstrates that a proper description of the NMD interaction requires including the finite nuclear size.
The graphs in Fig.~\ref{cs-muon} also demonstrate that muons are more sensitive to nuclear structure corrections than antimuons.

Overall, for (anti)muons the nuclear structure effects become very important already in intermediate relativistic regime and the scattering DCS differs fundamentally
from that of electrons and positrons. The nuclear structure effects (especially the NMD interaction)
are most pronounced at large scattering angles and larger collision velocities. This regime
corresponds to small impact parameters and large momentum
transfers, which effectively sample deeper regions
of the interaction potential, where the interaction between the projectile and the nucleus is the strongest. Due to this qualitatively distinct scattering dynamics
the (anti)muon scattering DCS can be a sensitive probe for both nuclear radii and magnetic
dipole moments.

%%%%%%%%%%%%%%%%%%%%%%%%%%%%%%%%%%%%%%%%%%%%%%%%%%%%%%%%%%%%%%%%%
\subsection{Polarization Changes}
The polarization of a scattered particle beam typically undergoes modification. Mott first predicted this effect for electrons in a Coulomb field \cite{mott29}, attributing it to the spin-orbit interaction. This can also be explained by noting that the upper and lower components of the Dirac wave function have different orbital angular momenta. Additionally, polarization changes can result from the interaction between incident particles and the angular momentum of the target system \cite{Burke1974,Vasileva2021PhysRevA.104.052808}. In this study, we consider the polarization change arising from the interaction of muons and antimuons with the NMD moment.

We assume that the momentum $\zhp$ of the incident muons or antimuons is along the $z$-axis, which is taken as the polar axis. Produced muon beams are usually strongly longitudinally polarized: muons are polarized along their momentum, while antimuons are polarized opposite to their momentum \cite{Blundell2021muon}. Therefore, the incident particle beam is taken to be $100\%$ polarized along the $z$-axis. The atomic nuclei are assumed to be initially unpolarized, as beams of polarized heavy nuclei are not yet available.

\begin{figure}[t]
\includegraphics[width=0.40\textwidth] {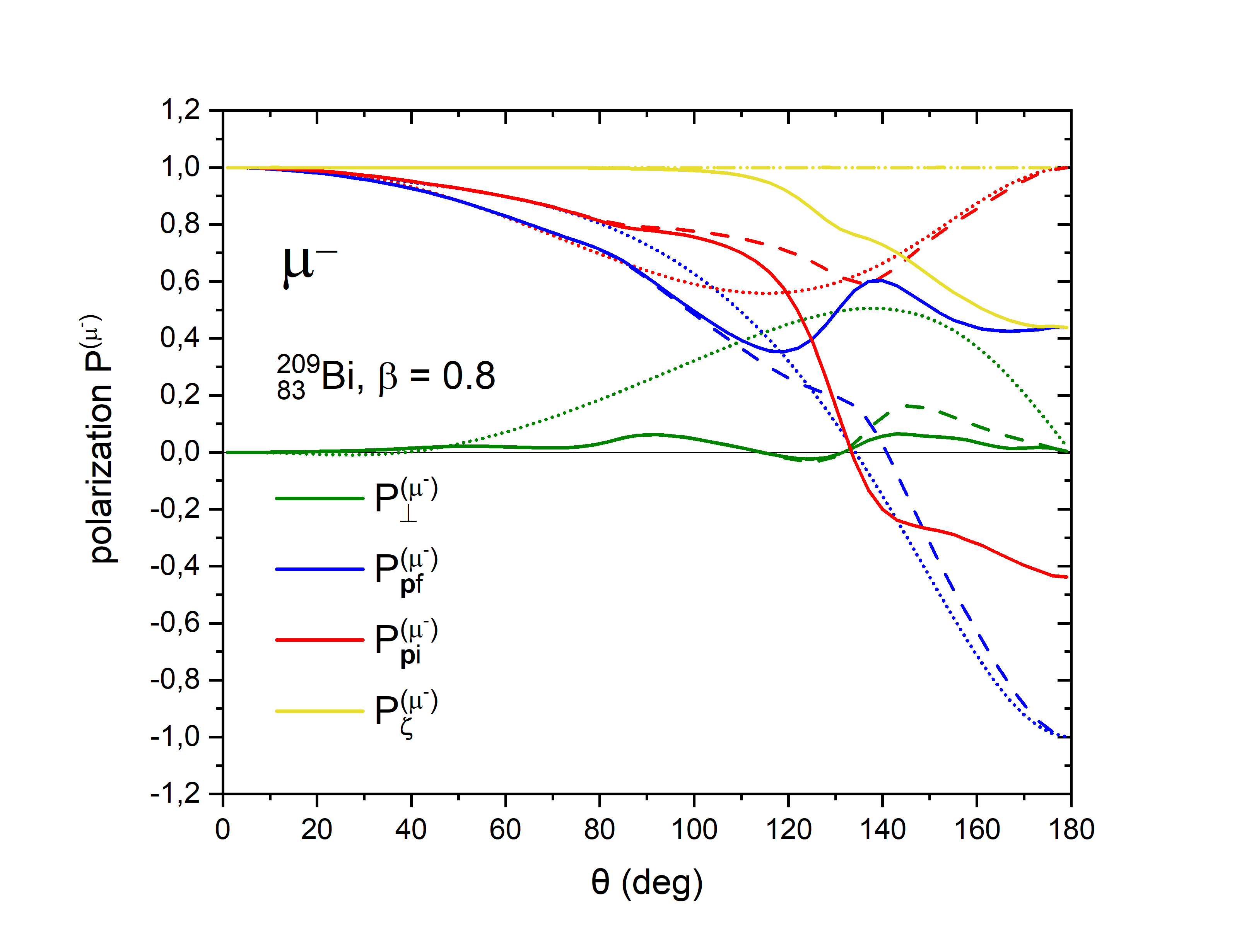}\
\includegraphics[width=0.40\textwidth] {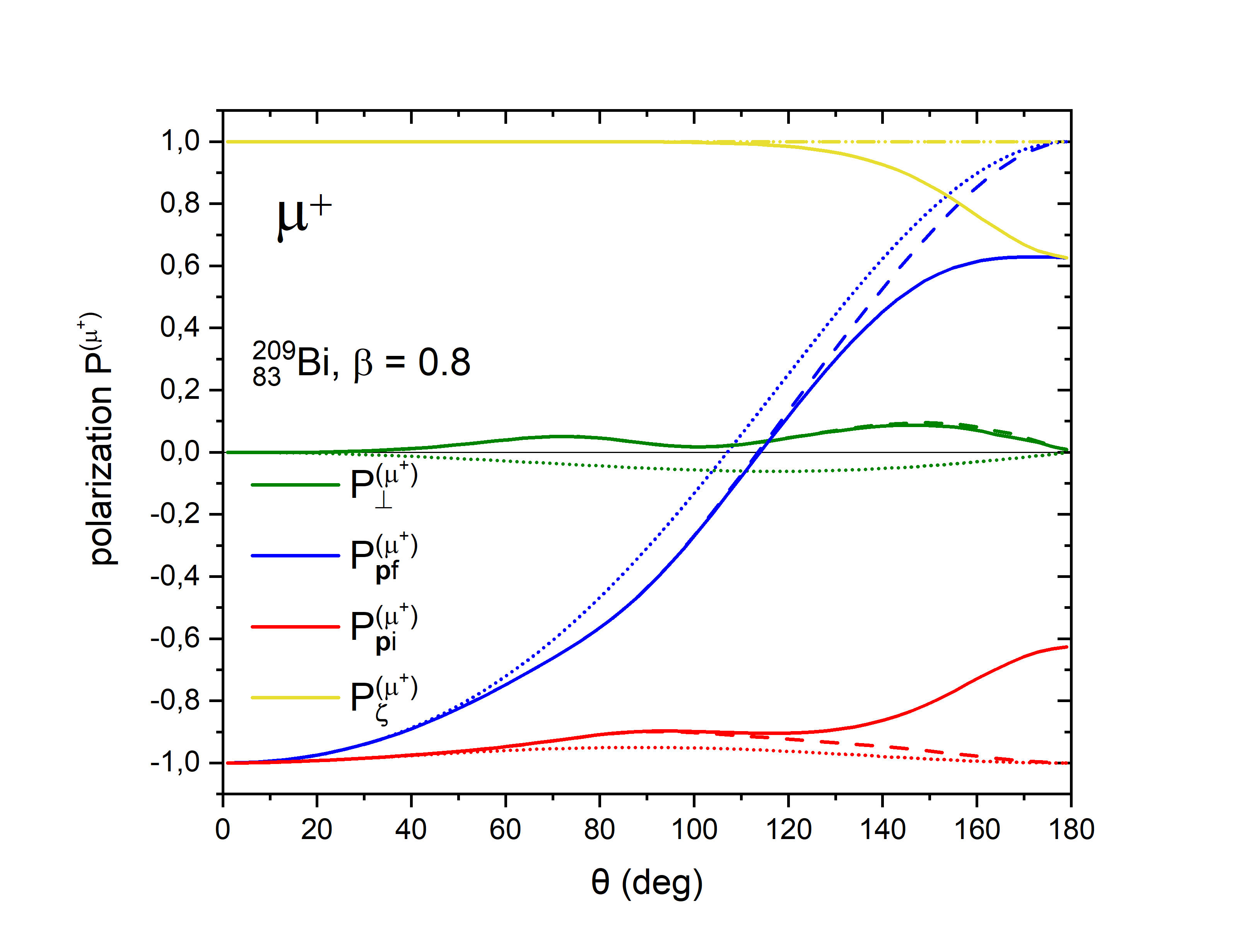}\
\caption{
Polarization of scattered muons (upper panel) and antimuons (lower panel) after scattering of $100\%$ polarized beams by unpolarized bare nuclei. The green, blue, and red solid curves represent the $P_{\perp}=P_{\zhe_{\perp}}$, $P_{\zhp_f}$, and $P_{\zhp_i}$ polarizations, respectively, as functions of the polar angle $\theta$. The yellow solid curves represent the degree of polarization. The dotted curves show the results obtained within the point-like nucleus approximation, while the dashed curves include NS corrections but no NMD correction.
}
\label{pol-muon}
\end{figure}

\begin{figure}[h!]
\includegraphics[width=0.40\textwidth] {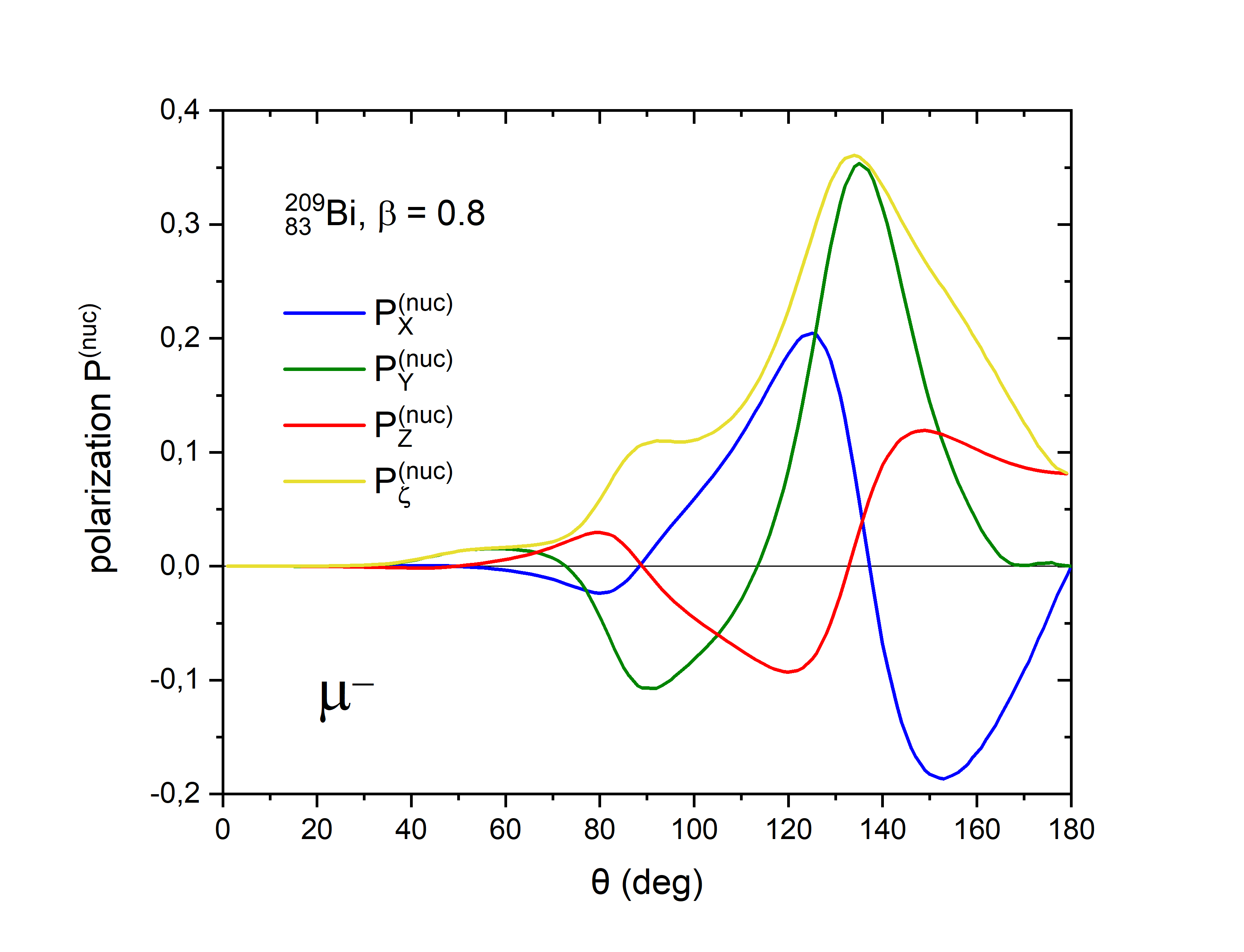}\
\includegraphics[width=0.40\textwidth] {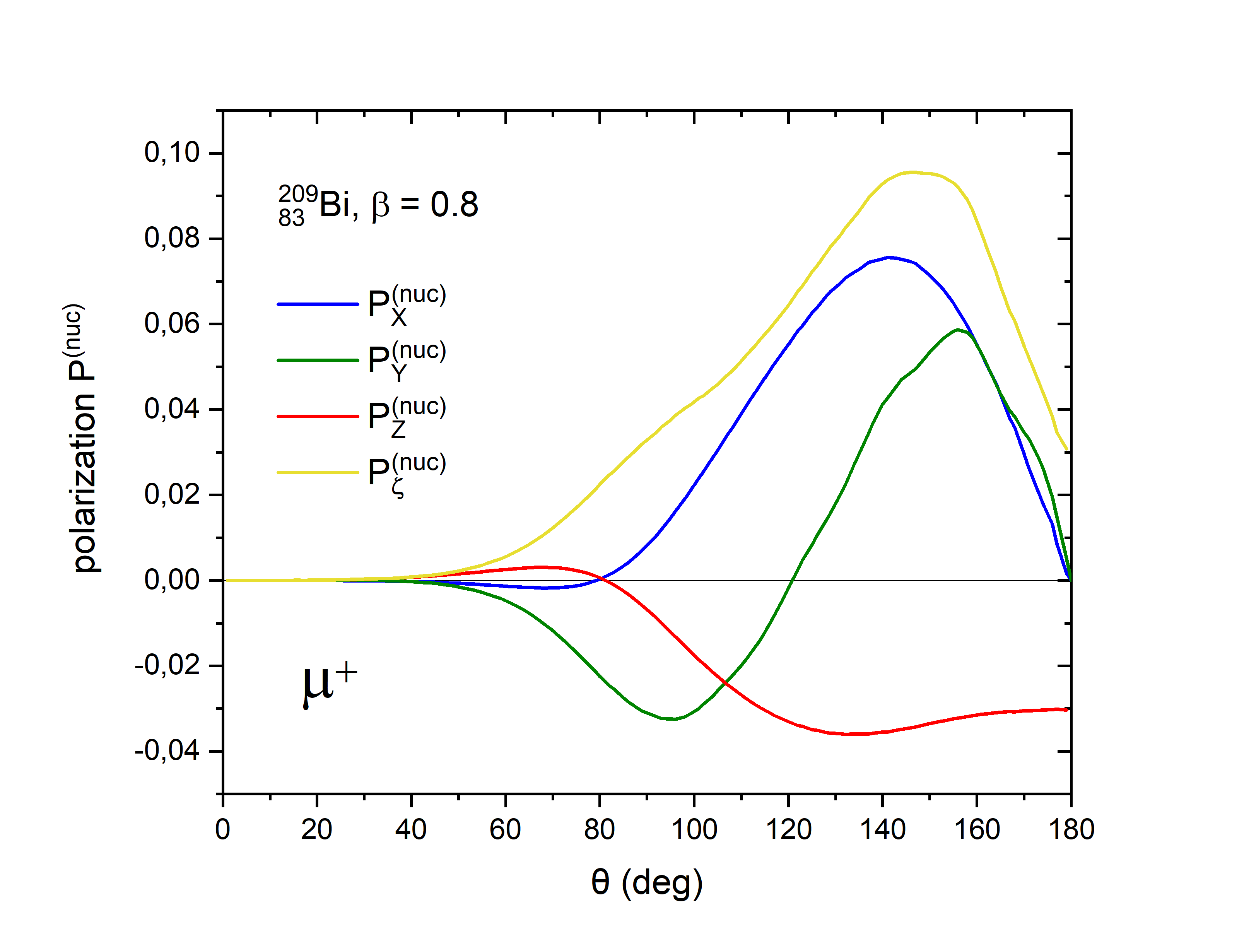}\
\caption{
Induced polarization of the atomic nucleus after scattering of $100\%$ polarized muons (upper panel) and antimuons (lower panel). The blue, green and red solid curves represent the $P^{(\txt{nuc})}_x=P^{(\txt{nuc})}_{\zhe_x}$, $P^{(\txt{nuc})}_y=P^{(\txt{nuc})}_{\zhe_{\perp}}$, and $P^{(\txt{nuc})}_z=P^{(\txt{nuc})}_{\zhp_i}$ 
the normalized average nuclear spin projections, respectively, as functions of the polar angle $\theta$. The yellow solid curves represent the polarization degree $P^{(\txt{nuc})}=|\zhzeta^{\txt{nuc}}_{\bar{S}}|$.
}
\label{pol-muon-nuc}
\end{figure}

The theoretical framework for describing the polarization of the lepton and the nucleus is presented in\ref{section-pol}.
The polarization after scattering for (anti-)muons and nuclei is described by the polarization vector $\zhzeta^{(l)}$ (see Eqs.~\Br{eqn260710n01}, \Br{polvec}) and normalized average nuclear spin vector $\zhzeta^\txt{nuc}_{\bar{S}}$ (\Eq{eqn260710n02}), respectively.
We are interested in the projections of these vectors on several specific directions naturally arising from the geometry of the scattering. Therefore we introduce the projection of the polarization vector $P_{\zhe}$ on a chosen axis $\zhe$
\begin{eqnarray}
P_{\zhe}
&=&
\frac{(\zhe,\zhzeta)}{|\zhe|}
\,,
\end{eqnarray}
which varies in the range $-1\le P_{\zhe}\le 1$. For the sake of convenience, we later refer to this quantity as the polarization on axis $\zhe$. Accordingly, the initial polarization $P^{(\txt{ini},\mu\pm)}_z$ is $1$ for the muon beam and $-1$ for the antimuon beam.

We also introduce an auxilliary unit vector perpendicular to the plane of scattering
\begin{eqnarray}
\zhe_{\perp}
&=&\label{eperp}
\frac{[\zhp_i\times\zhp_f]}{|[\zhp_i\times\zhp_f]|}
\,.
\end{eqnarray}
In Fig.~\ref{pol-muon}, we present the projections of scattered (anti)muon beam polarization vector $\zhzeta^{(\mu\pm)}$ for a collision velocity $\beta=0.8$ as functions of the polar angle $\theta$. The green curves show the $P_{\perp}=P_{\zhe_{\perp}}$ component. The blue curves show the $P_{\zhp_f}$ polarization on axis obtained by projecting the polarization vector onto the scattered muon momentum $\zhp_f$. The red curves show the $P_{\zhp_i}$ component obtained by projecting the polarization vector onto the incident muon momentum $\zhp_i$. The yellow solid curves represent the polarization degree $P^{(\mu\pm)}$. The dotted curves show results obtained within the point-like nucleus approximation, while the dashed curves include NS corrections but no NMD correction. The difference between the solid and dashed curves illustrates the role of the NMD corrections.

Scattering at small angles corresponds to a weak interaction between the scattered particle and the atomic nucleus. Accordingly, the projections of the polarization vector at $\theta=0^\circ$ coincide with the initial parameters of the incoming lepton beam.

We observe that the interaction of muons with the nuclear
magnetic moment drastically changes the polarization
of the scattered muon beam causing significant overall polarization loss at high scattering angles and, more importantly, shifting the polarization vector direction. This significantly alters the longitudinal polarization of the scattered muon beam. The polarization perpendicular to the scattering plane is less affected. The parameter describing this polarization component is commonly called the Sherman function, introduced for the electron scattering in \cite{sherman56}. The green dotted curves (the PN
results) can also be interpreted as the Sherman function
for electrons studied in \cite{sherman56}. Both the nuclear size effect and the NMD interaction reduce the Sherman function, resulting in a smaller perpendicular polarization change than for an electron beam.  We further
note that, for the parameters considered here, the
NMD interaction has a considerably larger influence on
muons than on antimuons.

In this work, we restrict our consideration to scattering by unpolarized nuclei. However, analyzing the DCS and the changes in muon polarization could also be utilized to diagnose or measure the initial polarization of a nuclear beam in future studies.

During the collision, both the muon (or antimuon) and the atomic nucleus undergo polarization changes. In Fig.~\ref{pol-muon-nuc}, we present the normalized average nuclear spin projections, which characterize the nuclear polarization acquired after the collision.  Specifically, we provide the projections of the nuclear polarization vector $\zhzeta^{\txt{nuc}}_{\bar{S}}$ \Eq{eqn260710n02} onto three perpendicular directions: $\zhe_x$, $\zhe_{\perp}$, and $\zhp_i$, where
\begin{eqnarray}
\zhe_x
&=&\label{ex}
\frac{[\zhe_{\perp}\times\zhp_f]}{|[\zhe_{\perp}\times\zhp_f]|}
\,.
\end{eqnarray}
We again conclude that the induced polarization after scattering of muons is much larger than that for antimuons.

It is worth noting that the nuclear polarization perpendicular to the scattering plane is remarkably high. This is particularly striking given that the initial state of the system possessed zero perpendicular polarization. Moreover, the longitudinal polarization (along the $z$-axis) is not dominant and remains comparable to the other projections. We attribute this behavior to the nature of the NMD interaction potential, which features the vector cross product of the target and lepton spin operators.

Fig.~\ref{pol-muon-nuc} shows the nuclear polarization change as a function
of the scattered muon angle, demonstrating
that nuclei can acquire significant polarization in the collision. The acquired polarization depends directly on the scattered muon momentum and is, therefore, correlated with the change in the momentum of the nucleus. If nuclei with specific transferred momenta can be separated from the primary beam, this process could serve as a method for producing polarized
nuclear beams, particularly for isotopes with large magnetic
dipole moments.

\begin{figure}[h!]
\includegraphics[width=0.4\textwidth] {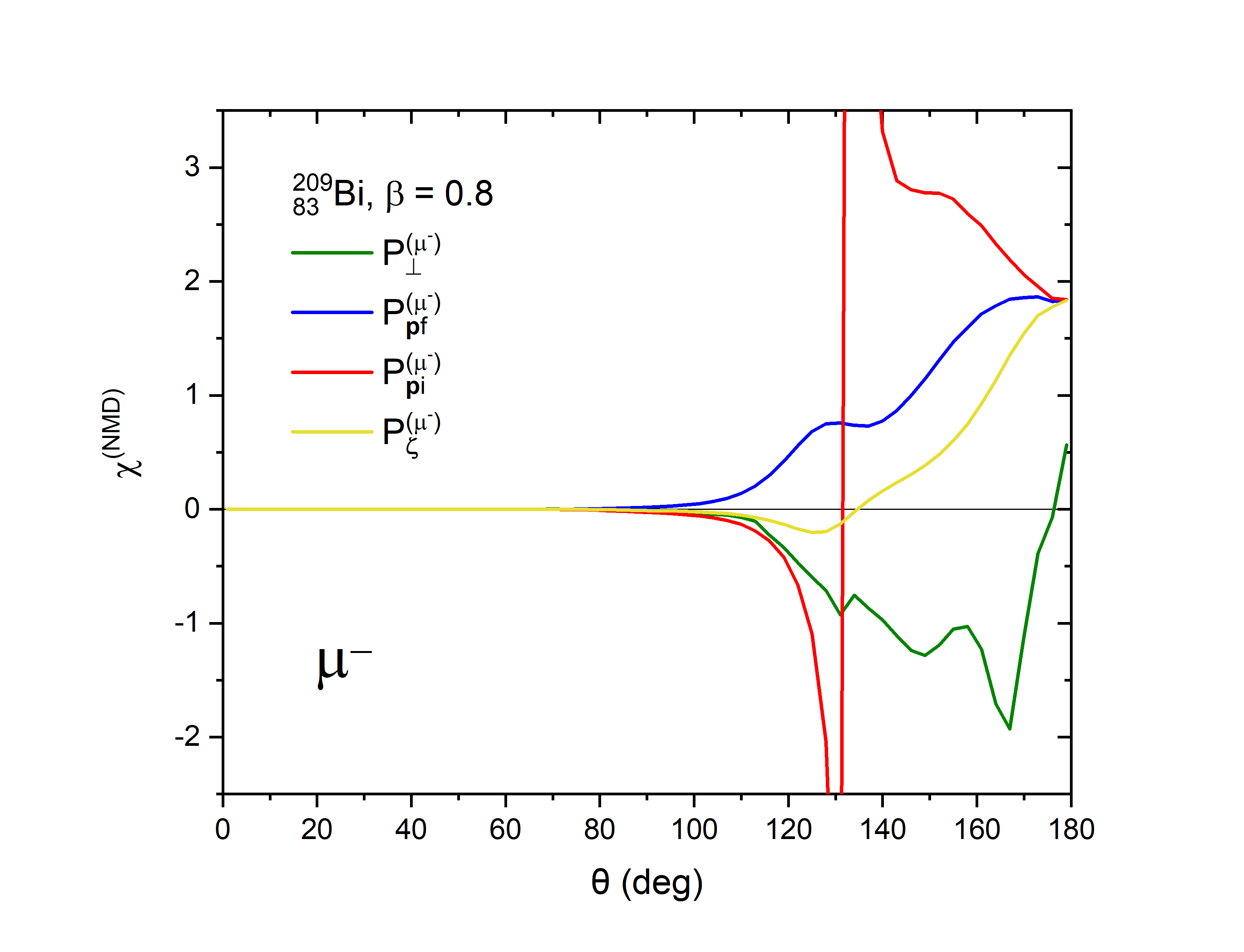}
\includegraphics[width=0.4\textwidth] {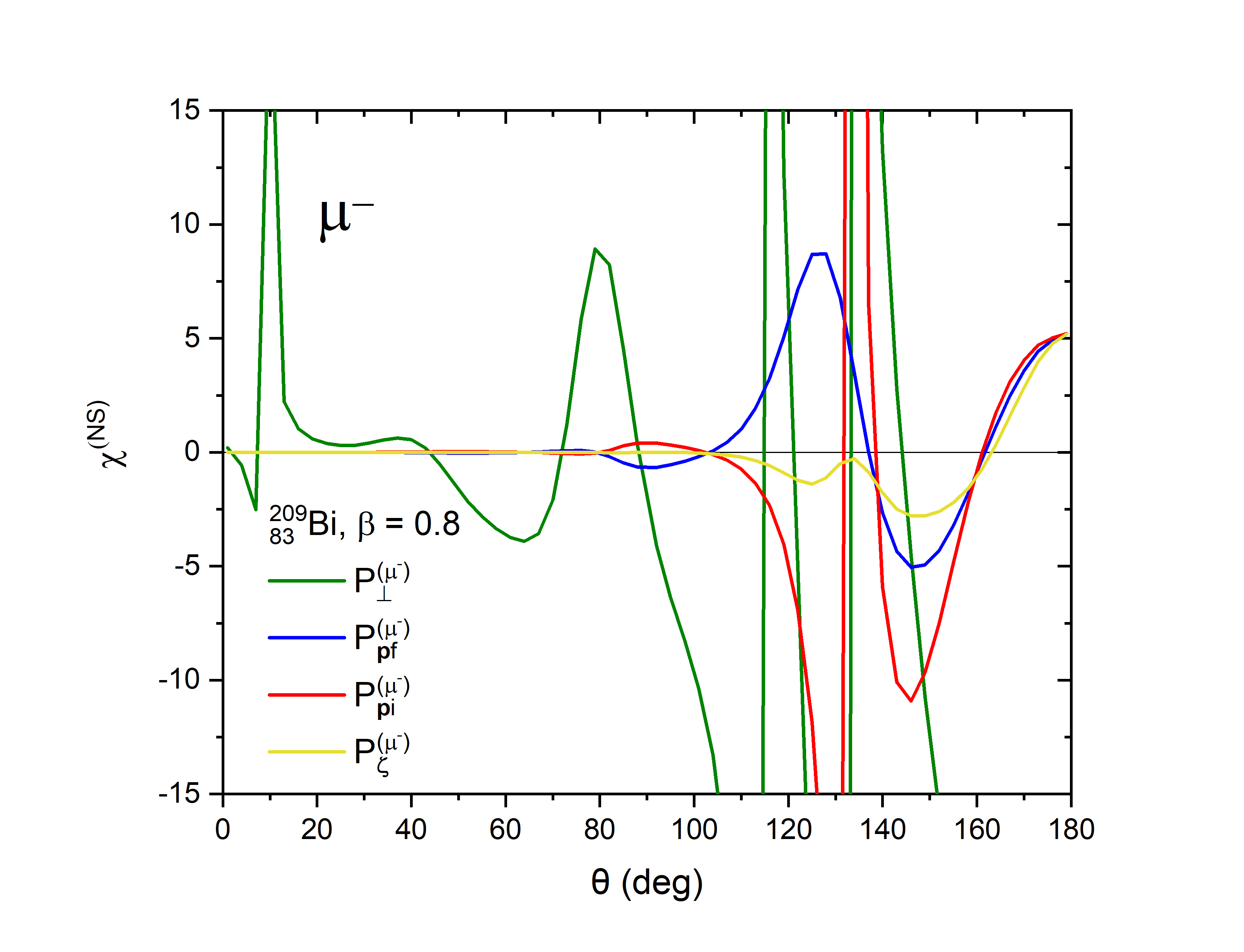}
\caption{Sensitivity ($\chi$) of the polarizations of scattered muons presented in Fig.~\ref{pol-muon} to the nuclear magnetic moment (upper panel) and to the root-mean-square charge radius of the nucleus (lower panel).}
\label{sens}
\end{figure}

The strong dependence of the polarization of scattered muons and antimuons on the nuclear magnetic dipole moment and the nuclear charge radius can be exploited to determine these parameters from polarization measurements. The corresponding sensitivities of the polarizations to these parameters can be expressed as
\begin{eqnarray}
\chi^{\txt{NMD}}
&=&\label{chi-nmd}
\frac{\mu}{P} \frac{dP}{d\mu}
\,,\quad
\chi^{\txt{NS}}
\,=\,\label{chi-ns}
\frac{r^{\txt{rms}}}{P} \frac{dP}{dr^{\txt{rms}}}
\,.
\end{eqnarray}
The sensitivity to the magnetic dipole moment is illustrated in the upper panel of Fig.~\ref{sens}. We present the function $\chi^{\txt{NMD}}$ corresponding to the characteristic projections $P$ of the polarization vector. Thus, if the polarization vector projection \(P\) is known with an accuracy of $1\%$, the magnetic dipole moment can be determined with an accuracy of ($1/\chi^{\txt{NMD}})\%$. Since certain polarization components can vanish at specific angles, the corresponding $\chi$ functions diverge at these points. In regions where the polarization is large, $\chi^{\txt{NMD}}$ can reach values of $1$--$2$.

The sensitivity of the polarizations to the nuclear root-mean-square charge radius, $\chi^{\txt{NS}}$, is presented in the lower panel of Fig.~\ref{sens}. In this case, $\chi^{\txt{NS}}$ can reach values as high as $10$--$20$.

Consequently, measurements of the polarization of scattered muons can be used to determine both the nuclear charge radius and the magnetic dipole moment. This approach presents an alternative to the widely used method of determining these parameters from the hyperfine structure in electronic atom spectroscopy or ultra-fast electron scattering.

\section{Conclusion}
In this work, we have demonstrated that nuclear structure corrections qualitatively alter the elastic scattering of muons and antimuons compared to that of electrons and positrons. Our analysis of the finite nuclear size effect and the nuclear magnetic dipole moment interaction for scattering by bare bismuth nuclei ($^{209}_{83}\text{Bi}$) yields several key findings. For electrons and positrons at collision velocities $\beta = 0.2$ to $0.8$, these corrections are negligible. In striking contrast, muons (and, to a lesser extent, antimuons) exhibit strong sensitivity. The differential cross section is modified by up to several orders of magnitude, especially at large scattering angles ($\theta > 90^\circ$), where the nuclear size correction alone suppresses the cross section by an order of magnitude.

While the nuclear size correction dominates, the nuclear magnetic dipole moment interaction is highly significant for nuclei with large magnetic moments, enhancing the differential cross section by a factor of several. Importantly, we find that the nuclear magnetic dipole moment contribution is itself strongly modulated by the finite nuclear size; a point-like approximation overestimates this effect, underscoring the necessity of treating both corrections simultaneously.

We have developed a theoretical approach to describe the polarization changes of both the scattered lepton beam and the atomic nucleus (assuming an initially unpolarized nuclear beam) for arbitrary nuclear spin. We have shown that these polarization changes are highly sensitive to both nuclear structure corrections. This sensitivity provides a practical pathway for determining the nuclear charge radius and the magnetic dipole moment from polarization measurements. Furthermore, this process can be utilized to measure the initial polarization of a nuclear beam.

\begin{acknowledgments}
The work of I.A.M. and K.N.L. was supported solely by the Russian Science Foundation under Grant No. 26-72-10004.
O.Y.A. was supported by the Chinese Academy of Sciences (CAS) Presidents International Fellowship Initiative (PIFI) under Grant Nos. 2025PVA0072.
\end{acknowledgments}

%%%%%%%%%%%%%%%%%%%%%%%%%%%%%%%%%%%%%%%%%%%%%%%%%%%%%%%%%%%%%%%%%%%%
\appendix
%%%%%%%%%%%%%%%%%%%%%%%%%%%%%%%%%%%%%%%%%%%%%%%%%%%%%%%%%%%%%%%%%%%%

%\appendix
%%%%%%%%%%%%%%%%%%%%%%%%%%%%%%%%%%%%%%%%%%%%%%%%%%%
\section{Furry-Sommerfeld-Maue approximation}
\label{appendix-fsm}

Matrix elements of the operator in \Eq{HFSnoIn02} evaluated with the functions in \Eq{psior} require careful treatment. Indeed, considering the matrix element of this operator with plane wave functions yields
\begin{eqnarray}
\int d^3\zhr e^{i\zhq\zhr} V^\txt{NMD}
&\sim&\label{eqn260602n06}
\int d^3\zhr \frac{e^{i\zhq\zhr}}{r^2}
\,\sim\,
\int^{\infty}_{0} dr \frac{e^{iqr}}{r}
\,,
\end{eqnarray}
where $\zhq$ is the momentum transfer (the difference between the initial and final electron momenta). Although this integral converges, it presents a significant challenge for numerical evaluation. Specifically, this difficulty manifests as slow convergence of the partial wave expansion \Eq{psior} when used for the lepton matrix element in \Eq{HFSnoIn02}. We note that the wave functions for the extended nucleus models are available only in the form of the partial wave expansion \Eq{psior}, with the number of terms limited to approximately $100$.

To overcome this issue, we exploit the fact that for a pure Coulomb potential $V=-\alpha Z/r$, the electron wave functions $\psi_{\dee \varkappa m}$ and the phases $\phi_{\varkappa}$ in \Eq{psior} can be derived analytically \cite{akhiezer65b}. Furthermore, within the Furry-Sommerfeld-Maue (FSM) approximation, the series in \Eq{psior} can be summed analytically to yield a closed-form expression \cite{Furry1934PhysRev.46.391,Sommerfeld1935,Bethe1954PhysRev.93.768,Elwert1969PhysRev.183.90}.

It is convenient to express the electron wave function ($\psi^\txt{C(\pm)}_{\zhp\mu}$) for a pure Coulomb potential as
\begin{eqnarray}
\psi^{\txt{C}(\pm)}_{\zhp\mu\zeta}(\zhr)
&=&\nonumber
\frac{(2\pi)^{3/2}}{\sqrt{p\varepsilon}} 
\\
&&\label{psiorc}
\hspace{-10pt}\times\sum_{\varkappa m}\Omega^{+}_{\varkappa m}(\zh\nu)\upsilon_{\mu}e^{i \phi^{\txt{C}(\pm)}_{\varkappa\zeta}}i^{l}\psi^{\txt{C}(\pm)}_{\varepsilon \varkappa m\zeta}(\zhr),
\end{eqnarray}
where
\begin{eqnarray}
\phi^{\txt{C}(+)}_{\varkappa\zeta}
&=&
-\frac{\pi\gamma_{\zeta}}{2}-\arg{\Gamma(\gamma_{\zeta}+i\xi)}
+\frac{\pi(l+1)}{2}
\,,
\\
\phi^{\txt{C}(-)}_{\varkappa\zeta}
&=&
-\phi^{\txt{C}(+)}_{\varkappa\zeta}
\,,
\\
\xi
&=&
\frac{\alpha Z \dee}{p}
\,,
\\
\gamma_{\zeta}
&=&\label{gammaz}
\sqrt{\varkappa^2-\zeta(\alpha Z)^2}
\,.
\end{eqnarray}
Here, we explicitly indicate the dependence on the parameter $\zeta$, which is initially set to 
\begin{eqnarray}
\zeta
&=&
1
\,.
\end{eqnarray}
The wave function $\psi^\txt{C(\pm)}_{\dee \varkappa m\zeta}$ can be written as
\begin{eqnarray}
\psi^\txt{C(\pm)}_{\dee \varkappa m\zeta}(\zhr)
&=&
\frac{1}{r}
\vectorb{g^\txt{C(\pm)}_{\dee \varkappa\zeta}(r)\Omega_{\varkappa m}(\zhn)}
{if^\txt{C(\pm)}_{\dee \varkappa\zeta}(r)\Omega_{\bar{\varkappa},m}(\zhn)}
\,.
\end{eqnarray}
We introduce the following parameters:
\begin{eqnarray}
\hat{\eta}^{(+)}_{\zeta}
&=&\label{eqn260602n02}
e^{i2\eta^{(+)}_{\zeta}}
\,=\,
\frac{-\varkappa+i\frac{\mx\xi}{\dee}}{\gamma_{\zeta}+i\xi}
\,,
\\
\hat{\eta}^{(-)}_{\zeta}
&=&\label{eqn260602n03}
e^{i2\eta^{(-)}_{\zeta}}
\,=\,
\frac{\gamma_{\zeta}-i\xi}{-\varkappa-i\frac{\mx\xi}{\dee}}
\,=\,
1/\hat{\eta}^{(+)*}_{\zeta}
\,.
\end{eqnarray}
We note that for $\zeta=1$, the following relations hold:
\begin{eqnarray}
|\hat{\eta}^{(\pm)}_{1}|
&=&\label{eqn260602n01}
1
\,,
\qquad
\hat{\eta}^{(+)}_{1}
\,=\,
\hat{\eta}^{(-)}_{1}
\,.
\end{eqnarray}
The radial functions $g^\txt{C}_{\dee \varkappa\zeta}$ and $f^\txt{C}_{\dee \varkappa\zeta}$ with the explicit parameter $\zeta$ read
\begin{eqnarray}
g^{\txt{C}(+)}_{\dee \varkappa\zeta}(r)
&=&
g^{}_{\dee \varkappa}(r,\gamma_{\zeta},\hat{\eta}^{(+)}_{\zeta})
\,,
\\
g^{\txt{C}(-)}_{\dee \varkappa\zeta}(r)
&=&
g_{\dee \varkappa}(r,\gamma_{\zeta},\hat{\eta}^{(-)}_{\zeta})/\hat{\eta}^{(-)}_{\zeta}
\,,
\\
%\end{eqnarray}
%\begin{eqnarray}
f^{\txt{C}(+)}_{\dee \varkappa\zeta}(r)
&=&
f^{}_{\dee \varkappa}(r,\gamma_{\zeta},\hat{\eta}^{(+)}_{\zeta})
\,,
\\
f^{\txt{C}(-)}_{\dee \varkappa\zeta}(r)
&=&
f_{\dee \varkappa}(r,\gamma_{\zeta},\hat{\eta}^{(-)}_{\zeta})/\hat{\eta}^{(-)}_{\zeta}
\,,
\end{eqnarray}
where
%\begin{widetext}
\begin{eqnarray}
g_{\dee \varkappa}(r,\gamma,\hat{\eta})
&=&\nonumber
\frac{1}{2}
e^{\frac{\pi\xi}{2}}
\frac{|\Gamma(\gamma+i\xi)|}{\Gamma(2\gamma+1)}
\sqrt{\frac{|\dee+\mx|}{\pi p}}
(2pr)^{\gamma}
\\
&&\nonumber
\times
\left[
\hat{\eta}%e^{i2\eta^{Z}_{\dee\varkappa}}
\left\{(\gamma+i\xi) e^{-ipr}
\right.\right.
\\
&&\nonumber
\times
\left.\left.
F(\gamma+1+i\xi,2\gamma+1,2ipr)\right\}\right.
\\
&&\nonumber
\phantom{e^{i2\eta}}
+
\left.
\left\{
(\gamma+i\xi)
e^{-ipr}
\right.\right.
\\
&&
\times
\left.\left.
F(\gamma+1+i\xi,2\gamma+1,2ipr)
\right\}^{*}
\right]
\,,
\\
f_{\dee \varkappa}(r,\gamma,\hat{\eta})
&=&\nonumber
i
\frac{1}{2}
e^{\frac{\pi\xi}{2}}
\frac{|\Gamma(\gamma+i\xi)|}{\Gamma(2\gamma+1)}
\sqrt{
\frac{|\dee-\mx|}{\pi p}}\frac{\dee}{|\dee|}
(2pr)^{\gamma}
\\
&&\nonumber
\times
\left[
\hat{\eta}%e^{i2\eta^{Z}_{\dee\varkappa}}
\left\{(\gamma+i\xi) e^{-ipr}
\right.\right.
\\
&&\nonumber
\left.\left.
\times
F(\gamma+1+i\xi,2\gamma+1,2ipr)\right\}\right.
\\
&&\nonumber
\phantom{e^{i2\eta}}
-
\left.
\left\{
(\gamma+i\xi)
e^{-ipr}
\right.\right.
\\
&&
\left.\left.
\times
F(\gamma+1+i\xi,2\gamma+1,2ipr)
\right\}^{*}
\right]
\,.
\end{eqnarray}

Within the FSM approximation, the wave function \Eq{psiorc} is replaced by the following closed-form expression \cite{Furry1934PhysRev.46.391,Sommerfeld1935,Bethe1954PhysRev.93.768,Elwert1969PhysRev.183.90}:
\begin{eqnarray}
\Phi^{\txt{FSM}(\pm)}_{\zhp\mu}(\zhr)
&=&\nonumber
\Gamma(1\mp i\xi)e^{\pi\xi/2} e^{i\zhp\zhr}
\left(1-\frac{i\alpha\zhnabla}{2\dee}\right)
\\
&&\label{eqn260602n20}
\hspace{-10pt}
\times
F(\pm i\xi,1,-i(\zhp\zhr\mp pr)) v^{\mu}(\zhp)
\,.
\end{eqnarray}
The FSM approximation can be formally obtained from the wave function \Eq{psiorc} by setting the parameter $\zeta$ to zero ($\zeta=0$). We note that for $\zeta=0$, the relations in \Eq{eqn260602n01} no longer hold. Consequently, the parameters $\eta^{(\pm)}_0$ defined in Eqs.~\Br{eqn260602n02} and \Br{eqn260602n03} become complex values. The connection between the exact Coulomb-Dirac solution \Eq{psiorc} and the FSM approximation was discussed in \cite{Bethe1954PhysRev.93.768,Jakubassa2013}. Thus, we can write
\begin{eqnarray}
\Phi^{\txt{FSM}(\pm)}_{\zhp\mu}(\zhr)
&=&\label{eqn260602n05}
\psi^{\txt{C}(\pm)}_{\zhp\mu 0}(\zhr)
\,.
\end{eqnarray}
This identity has been verified numerically using various matrix elements.

From \Eq{gammaz}, it follows that $\gamma_1$ and $\gamma_0$ converge toward each other as $|\varkappa|$ increases. Accordingly, the partial wave expansion \Eq{psiorc} for the difference $\psi^{\txt{C}(\pm)}_{\zhp\mu1}-\psi^{\txt{C}(\pm)}_{\zhp\mu 0}$ converges rapidly with respect to $|\varkappa|$. Furthermore, because the nuclear potentials for the point-like model and the extended nucleus model differ only in the immediate vicinity of the nucleus, the partial wave expansion for the difference $\psi^{(\pm)}_{\zhp\mu}-\psi^{\txt{C}(\pm)}_{\zhp\mu 0}$ is also rapidly convergent.

Using these notations, we can address the calculation of the integral in \Eq{HFSnoIn02}. The exact electron wave function can be decomposed as
\begin{eqnarray}
\psi^{(\pm)}_{\zhp\mu}
&=&
\Phi^{\txt{FSM}(\pm)}_{\zhp\mu}
+\left(\psi^{(\pm)}_{\zhp\mu} - \Phi^{\txt{FSM}(\pm)}_{\zhp\mu}\right)
\,.
\end{eqnarray}
Using the identity \Eq{eqn260602n05}, a matrix element can be written as
\begin{eqnarray}
\langle\psi^{(-)}_{\zhp_f\mu_f}|V|\psi^{(+)}_{\zhp_i\mu_i}\rangle
&=&\label{eqn260602n11}
\langle\Phi^{\txt{FSM}(-)}_{\zhp_f\mu_f}|V|\Phi^{\txt{FSM}(+)}_{\zhp_i\mu_i}\rangle
\\
&&\label{eqn260602n12}
\hspace{-80pt}
+\langle(\psi^{(-)}_{\zhp_f\mu_f}-\psi^{\txt{C}(-)}_{\zhp_f\mu_f0})|V|\psi^{\txt{C}(+)}_{\zhp_i\mu_i0}\rangle
\\
&&\label{eqn260602n13}
\hspace{-80pt}
+\langle\psi^{\txt{C}(-)}_{\zhp_f\mu_f0}|V|(\psi^{(+)}_{\zhp_i\mu_i}-\psi^{\txt{C}(+)}_{\zhp_i\mu_i0})\rangle
\\
&&\label{eqn260602n14}
\hspace{-80pt}
+\langle(\psi^{(-)}_{\zhp_f\mu_f}-\psi^{\txt{C}(-)}_{\zhp_f\mu_f0})|V|(\psi^{(+)}_{\zhp_i\mu_i}-\psi^{\txt{C}(+)}_{\zhp_i\mu_i0})\rangle
\,.
\end{eqnarray}
In this decomposition, the terms \Br{eqn260602n12}--\Br{eqn260602n14} are calculated using the partial wave expansions as in Eqs.~\Br{psior} and \Br{psiorc}, which converge rapidly. The primary term \Br{eqn260602n11} is computed numerically with the use of the closed-form expression in \Eq{eqn260602n20} as a full three-dimensional integral without relying on a partial wave expansion.

%%%%%%%%%%%%%%%%%%%%%%%%%%%%%%%%%%%%%%%%%%%%%%%%%%%%%%%%%%%
\section{Partial-wave expansion of wave functions in NMD matrix elements}
\label{appendix-angle}
The correction to the scattering amplitude arising from the interaction of the scattering lepton with the nuclear magnetic dipole (NMD) is given by Eqs.~\Br{VNMD}, \Br{mU}. By expressing the scalar product in terms of spherical components, the amplitude can be decomposed into nuclear and leptonic parts \Br{HFSnoIn02}:
\begin{eqnarray}
\Delta U^{\txt{NMD}}
&=&\nonumber
\langle
\psi^{}_{IM_f,\dee\zhp_f\mu_f}|
\Delta V^{\txt{NMD}}
|\psi^{}_{IM_i,\dee\zhp_i\mu_i}\rangle
\\
&=&
-\frac{q\mu\mu_N}{I}
\sum\limits_{\eta=\pm1,0} (-1)^{\eta}
N_{\bar{\eta}}L_{\eta}
\,,
\end{eqnarray}
where
\begin{eqnarray}
N_{\eta}
&=&
\langle
\psi^{(\txt{nuc})}_{IM_f}|
\hI_{\mu}
|\psi^{(\txt{nuc})}_{IM_i}\rangle
\,,
\\
L_{\eta}
&=&
\langle
\psi^{(-)(\ell)}_{\dee,\zhp_f\mu_f}|
\frac{[\zhr\times\zhalpha]_{\eta}}{r^3}
|\psi^{(+)(\ell)}_{\dee,\zhp_i\mu_i}\rangle
\,.
\end{eqnarray}

The nuclear matrix element can be expressed in terms of a Clebsch-Gordan coefficient \cite{Varshalovich1988QuantumTO}
\begin{eqnarray}
N_{\eta}
&=&
\sqrt{I(I+1)}\,C^{IM_f}_{IM_i1\mu}
\,.
\end{eqnarray}

Using the partial-wave expansions for the incoming and outgoing lepton wave functions, as given in \Eq{psior}, we obtain
\begin{eqnarray}
L_{\eta}
&=&\nonumber
\frac{(2\pi)^3}{p\dee}
\sum_{j_fl_fm_f}\sum_{j_il_im_i}
e^{i(\phi_{j_fl_f}+\phi_{j_il_i})}i^{-l_f+l_i}
\\
&&\nonumber
\times
\left(\Omega^+_{j_fl_fm_f}(\zhnu_f)v_{\mu_f}(\zhnu_f)\right)^{*}
\\
&&\nonumber
\times
\left(\Omega^+_{j_il_im_i}(\zhnu_i)v_{\mu_i}(\zhnu_i)\right)
\\
&&\label{eqn260703n01}
\times
\langle\psi^{(\ell)}_{\dee,j_fl_fm_f}|
\frac{[\zhr\times\zhalpha]_{\eta}}{r^3}
|\psi^{(\ell)}_{\dee,j_il_im_i}\rangle
\,,
\end{eqnarray}
where
\begin{eqnarray}
\psi^{(\ell)}_{\varepsilon, jlm}(\zhr)
&=&\frac{1}{r}
\left(
\begin{array}{cc}
g_{jl}(r)\Omega_{jlm}(\zh n)\\
if_{jl}(r)\Omega_{j,2j-l,m}(\zh n)\\
\end{array}
\right)
\,.
\end{eqnarray}

Taking the incident lepton direction to be along the $z$-axis, i.e. $\zhnu_i=\zhe_z$, we can write
\begin{eqnarray}
\left(\Omega^+_{j_il_im_i}(\zhnu_i)v_{\mu_i}(\zhnu_i)\right)
&=&
\frac{2l_i+1}{4\pi}
C^{j_i\mu_i}_{l_i 0,\frac{1}{2}\mu_i}
%\langle l_i0\frac{1}{2}\mu_i|j_i\mu_i\rangle
\,.
\end{eqnarray}

We now consider the last matrix element appearing in \Eq{eqn260703n01}
\begin{eqnarray}
D
&=&
\langle\psi^{(\ell)}_{\dee,j_fl_fm_f}|
\frac{[\zhr\times\zhalpha]_{\eta}}{r^3}
|\psi^{(\ell)}_{\dee,j_il_im_i}\rangle
\,.
\end{eqnarray}
The integration over the angular and spinor variables can be performed analytically, yielding
\begin{eqnarray}
D
&=&
\sqrt{\frac{8\pi}{3}}
\xi(l_f,\xl_i)
\int dr
\frac{1}{r^2}
\left(g^{*}_ff_i 
+f^{*}_f g_i\right)
\,,
\end{eqnarray}
where $l'_{i,f}=2j_{i,f}-l_{i,f}$, and $g_{i,f}$ and $f_{i,f}$ are the upper and lower radial components of the Dirac wave functions for the initial and final states, respectively.
The angular coefficient $\xi(\ell_f,\ell_i)$ is given by
\begin{eqnarray}
\xi(\ell_f,\ell_i)
&=&\nonumber
(-1)^{1+j_i+\ell_f-m_f}
\sqrt{\frac{3}{4\pi}}
\threej{j_f}{1}{j_i}{\xm_f}{\mu}{m_i}
\\
&&\nonumber
\times
%\Pi(1,j_f,j_i)
\sqrt{(2j_f+1)(2j_i+1)}
\threej{j_i}{j_f}{1}{\frac{1}{2}}{\frac{1}{2}}{-1}
\\
&&
\times
\frac{1}{2}\left(1-(-1)^{\ell_f+\ell_i}\right)
\,.
\end{eqnarray}

%%%%%%%%%%%%%%%%%%%%%%%%%%%%%%%%%%%%%%%%%%%%%%%%%%%%%%%%%%%%
\section{Polarization theory for elastic scattering of leptons by targets with arbitrary $S$}
\label{appendixpol}
In this appendix we discuss a generalization for the description of the polarization change of the lepton that occurs in an elastic scattering process involving spin-spin interaction. The incident particle is assumed to have spin of $1/2$, while the target may carry an arbitrary spin $S$. A distinctive feature of elastic scattering is that it is invariant under time reversal, which puts restrictions on the scattering matrix $\hat{M}$. 

Since the scattering matrix $\hat{M}$ describes the polarization change of two particles, it is convenient to represent $\hat{M}$ as an expansion over direct products of two operators. These operators belong to two independent basis sets, one associated with the incident particle and the other with the target:
\begin{eqnarray}
	\hat{M}&=&\sum_{i=1}^{(2S+1)^2}\sum_{j=1}^{4}a_{ij}\hat{T}_{i}(S)\otimes\hat{T}_{j}(1/2)\,. \label{M_exp_gen}
\end{eqnarray}

For particles with spin of $1/2$ the conventional basis set consists of $2 \times 2$ identity matrix $\hat{I}_2$ and three Pauli matrices $\hat{\zhsigma}$. In order to take into account inherent symmetries of the elastic scattering it is convenient to choose new orthogonal basis vectors $\zhn$, $\zhk$, $\zhq$ that transform under time reversal $T$ and parity $P$ as follows:
\begin{eqnarray}
	\zhn=\dfrac{\zhp_i \times \zhp_f}{|\zhp_i \times \zhp_f|}, \quad &\zhn\xrightarrow{T}-\zhn, \quad &\zhn\xrightarrow{P}\zhn
\,,
\\
	\zhk=\dfrac{\zhp_i + \zhp_f}{|\zhp_i + \zhp_f|}, \quad &\zhk\xrightarrow{T}-\zhk, \quad &\zhk\xrightarrow{P}-\zhk
\,,
\\
	\zhq=\dfrac{\zhp_i - \zhp_f}{|\zhp_i - \zhp_f|}, \quad &\zhq\xrightarrow{T}\zhq, \quad &\zhq\xrightarrow{P}-\zhq
\,.
\end{eqnarray}
Under time reversal $T$ and parity $P$ any spin operator transforms as
\begin{eqnarray}
	&\zhsigma\xrightarrow{T}-\zhsigma, \quad &\zhsigma\xrightarrow{P}\zhsigma
\,,
\\
	&\zhS\xrightarrow{T}-\zhS, \quad &\zhS\xrightarrow{P}\zhS
  \,.
\end{eqnarray}
With this basis, the spin-$1/2$ operators fall into three symmetry types under $P$ and $T$: those transforming as $\zhsigma \zhn$, $\zhsigma \zhk$ and $\zhsigma \zhq$.
Since matrix $\hat{M}$ is invariant under both $P$ and $T$, expansion \Eq{M_exp_gen} contains only direct products of operators with matching symmetry types:
\begin{eqnarray}
	\hat{M}&=&a_0 \hat{I}_{2S+1}\otimes \hat{I}_2+a_n^{(1)} \hat{I}_{2S+1}\otimes(\hat{\zhsigma}\zhn)\nonumber\\
	&+&\hat{A}_n^{(2)} \otimes \hat{I}_2 +\hat{A}_n^{(12)} \otimes (\hat{\zhsigma}\zhn)\nonumber\\
	&+&\hat{A}_k^{(12)} \otimes (\hat{\zhsigma}\zhk)+\hat{A}_q^{(12)} \otimes (\hat{\zhsigma}\zhq)\,, \label{scatmatexp}
\end{eqnarray}
where $a_i$ are numerical coefficients and $\hat{A}_i$ are $(2S+1) \times (2S+1)$ operators with subscript $i$ indicating symmetry type. The operators $\hat{A}_i$ can be constructed as sums of terms such as $(\hat{\zhS}\zhn)^\alpha (\hat{\zhS}\zhk)^\beta (\hat{\zhS}\zhq)^\gamma$ with $\alpha+\beta+\gamma \leqslant 2S$ \cite{Varshalovich1988QuantumTO}, where the symmetry type is determined by the following conditions: for $\hat{A}_n$, $\beta$ and $\gamma$ are even; for $\hat{A}_k$, $\beta$ is odd and $\gamma$ is even; for $\hat{A}_q$, $\beta$ is even and $\gamma$ is odd.

The structure of the scattering matrix $\hat{M}$ in \Eq{scatmatexp} determines the possible polarization changes during scattering. To extract the polarization parameters relevant to the experiment, we need traces of the form $\trace(\hat{M}\hat{\rho}_i\hat{M}^\dagger\hat{\rho}_f)$. Using $\trace(A \otimes B)=\trace A \trace B$, such traces become sums of products of separate spin-$1/2$ particle and target parts. The traces of form $\trace [(\hat{\zhS} \zhn)^\alpha(\hat{\zhS} \zhk)^\beta(\hat{\zhS} \zhq)^\gamma]$ are identically zero unless $\alpha,\beta,\gamma$ are simultaneously even or simultaneously odd \cite{Subramanian_1974}, therefore
\begin{eqnarray}
	\trace \hat{A}_k&=&\trace \hat{A}_q\,=\,\trace [\hat{A}_k \hat{A}_n]\,=\,\trace [\hat{A}_q \hat{A}_n]\,=\,0\,.
\end{eqnarray}
Assuming that the target is initially unpolarized and the polarization of the incident beam is described by polarization vector $\zhzeta_i$, we obtain the following formulae for the cross section $\frac{d\sigma}{d\Omega}$ and the polarization $\zhzeta_f$ of the spin-$1/2$ particle beam after scattering:
\begin{eqnarray}
	\dfrac{d\sigma}{d\Omega}&=&\dfrac{\varepsilon_i^2}{(2\pi)^2}\left\{I(\theta)+(\zhn \zhzeta_i)\left[D(\theta)+\Delta(\theta)\right]\right\}
\,;
\end{eqnarray}
\begin{eqnarray}
	\zhzeta_f&=&\dfrac{D(\theta)-\Delta(\theta)}{\tilde{I}_{\zhzeta_i}(\theta)}\zhn\nonumber\\
	&+&	\left(\dfrac{I(\theta)}{\tilde{I}_{\zhzeta_i}(\theta)}-\dfrac{H(\theta)+K(\theta)}{\tilde{I}_{\zhzeta_i}(\theta)}\right)(\zhn\zhzeta_i)\zhn\nonumber\\
	&+&	\left(\dfrac{I(\theta)}{\tilde{I}_{\zhzeta_i}(\theta)}-\dfrac{G(\theta)+K(\theta)}{\tilde{I}_{\zhzeta_i}(\theta)}\right)(\zhk\zhzeta_i)\zhk\nonumber\\
	&+&	\left(\dfrac{I(\theta)}{\tilde{I}_{\zhzeta_i}(\theta)}-\dfrac{H(\theta)+G(\theta)}{\tilde{I}_{\zhzeta_i}(\theta)}\right)(\zhq\zhzeta_i)\zhq\nonumber\\
	&+&\left(\dfrac{L(\theta)+F(\theta)}{\tilde{I}_{\zhzeta_i}(\theta)}\right)(\zhq\zhzeta_i)\zhk\nonumber\\
	&+&\left(\dfrac{L(\theta)-F(\theta)}{\tilde{I}_{\zhzeta_i}(\theta)}\right)(\zhk\zhzeta_i)\zhq
\,,
\end{eqnarray}
where
\begin{eqnarray}
	I(\theta)
        &=&\nonumber
        \dfrac{1}{2}\dfrac{1}{2S+1}\left\{ \trace \left[(a_0+\hat{A}_n^{(2)})(a_0+\hat{A}_n^{(2)})^\dagger\right] \right.\\
	 &&\nonumber
          +\trace \left[(a_n^{(1)}+\hat{A}_n^{(12)})(a_n^{(1)}+\hat{A}_n^{(12)})^\dagger\right]
         \\
	 &&\nonumber
          + \left.\trace \left[\hat{A}_k^{(12)}(\hat{A}_k^{(12)})^\dagger\right]\right.
          \\
          &&
          + \left.\trace \left[ \hat{A}_q^{(12)}(\hat{A}_q^{(12)})^\dagger\right] \right\}
\,,
\end{eqnarray}
\begin{eqnarray}
	\tilde{I}_{\zhzeta_i}(\theta)
        &=&
        I(\theta)+(\zhn \zhzeta_i)\left[D(\theta)+\Delta(\theta)\right]
,
\end{eqnarray}
\begin{eqnarray}
	D(\theta)
        &=&
        \dfrac{1}{2S+1}\nonumber
        \\
        &&
        \!\!\!\!\!\!\!\!\times
        \real\left\{\trace\left[(a_n^{(1)}+\hat{A}_n^{(12)})(a_0+\hat{A}_n^{(2)})^\dagger\right]\right\}
,
\end{eqnarray}
\begin{eqnarray}
	\Delta(\theta)&=&\dfrac{1}{2S+1}\image\left\{\trace\left[ (\hat{A}_k^{(12)})^\dagger\hat{A}_q^{(12)}\right]\right\}
,
\end{eqnarray}
\begin{eqnarray}
	F(\theta)
        &=&\nonumber
        \dfrac{1}{2S+1}
        \\
        &&
        \!\!\!\!\!\!\!\!\times
        \image\left\{ \trace\left[(a_0+\hat{A}_n^{(2)})(a_n^{(1)}+\hat{A}_n^{(12)})^\dagger\right]\right\}
,
\end{eqnarray}
\begin{eqnarray}
	G(\theta)
        &=&\nonumber
        \dfrac{1}{2S+1}
        \\
        &&
        \times
        \trace\left[(a_n^{(1)}+\hat{A}_n^{(12)})(a_n^{(1)}+\hat{A}_n^{(12)})^\dagger\right]
,
\end{eqnarray}
\begin{eqnarray}
	H(\theta)&=&\dfrac{1}{2S+1} \trace \left[\hat{A}_k^{(12)}(\hat{A}_k^{(12)})^\dagger\right]
,
\end{eqnarray}
\begin{eqnarray}
	K(\theta)&=&\dfrac{1}{2S+1} \trace \left[\hat{A}_q^{(12)}(\hat{A}_q^{(12)})^\dagger\right]
,
\end{eqnarray}
\begin{eqnarray}
	L(\theta)&=&\dfrac{1}{2S+1} \real\left\{\trace \left[ \hat{A}_k^{(12)}(\hat{A}_q^{(12)})^\dagger\right]\right\}.
\end{eqnarray}
The traces above are evaluated over matrices of size $2(2S+1) \times 2(2S+1)$. That is, an implicit identity matrix of the appropriate dimension is assumed within each trace.

For spin-$1/2$ targets, parameters $L(\theta)$ and $\Delta(\theta)$ are equal to zero. Notably, for targets with spin $S>1/2$, the asymmetry function (azimuthal cross‑section dependence) differs from the Sherman function (acquired polarization in an initially unpolarized system). Moreover, in‑plane polarization changes are less restricted.


\begin{thebibliography}{48}%
\makeatletter
\providecommand \@ifxundefined [1]{%
 \@ifx{#1\undefined}
}%
\providecommand \@ifnum [1]{%
 \ifnum #1\expandafter \@firstoftwo
 \else \expandafter \@secondoftwo
 \fi
}%
\providecommand \@ifx [1]{%
 \ifx #1\expandafter \@firstoftwo
 \else \expandafter \@secondoftwo
 \fi
}%
\providecommand \natexlab [1]{#1}%
\providecommand \enquote  [1]{``#1''}%
\providecommand \bibnamefont  [1]{#1}%
\providecommand \bibfnamefont [1]{#1}%
\providecommand \citenamefont [1]{#1}%
\providecommand \href@noop [0]{\@secondoftwo}%
\providecommand \href [0]{\begingroup \@sanitize@url \@href}%
\providecommand \@href[1]{\@@startlink{#1}\@@href}%
\providecommand \@@href[1]{\endgroup#1\@@endlink}%
\providecommand \@sanitize@url [0]{\catcode `\\12\catcode `\$12\catcode
  `\&12\catcode `\#12\catcode `\^12\catcode `\_12\catcode `\%12\relax}%
\providecommand \@@startlink[1]{}%
\providecommand \@@endlink[0]{}%
\providecommand \url  [0]{\begingroup\@sanitize@url \@url }%
\providecommand \@url [1]{\endgroup\@href {#1}{\urlprefix }}%
\providecommand \urlprefix  [0]{URL }%
\providecommand \Eprint [0]{\href }%
\providecommand \doibase [0]{https://doi.org/}%
\providecommand \selectlanguage [0]{\@gobble}%
\providecommand \bibinfo  [0]{\@secondoftwo}%
\providecommand \bibfield  [0]{\@secondoftwo}%
\providecommand \translation [1]{[#1]}%
\providecommand \BibitemOpen [0]{}%
\providecommand \bibitemStop [0]{}%
\providecommand \bibitemNoStop [0]{.\EOS\space}%
\providecommand \EOS [0]{\spacefactor3000\relax}%
\providecommand \BibitemShut  [1]{\csname bibitem#1\endcsname}%
\let\auto@bib@innerbib\@empty
%</preamble>
\bibitem [{\citenamefont {Walecka}(2023)}]{Walecka_2023}%
  \BibitemOpen
  \bibfield  {author} {\bibinfo {author} {\bibfnamefont {J.~D.}\ \bibnamefont
  {Walecka}},\ }\href@noop {} {\emph {\bibinfo {title} {Electron Scattering for
  Nuclear and Nucleon Structure}}},\ Cambridge Monographs on Particle Physics,
  Nuclear Physics and Cosmology\ (\bibinfo  {publisher} {Cambridge University
  Press},\ \bibinfo {year} {2023})\BibitemShut {NoStop}%
\bibitem [{\citenamefont {Andreev}\ \emph {et~al.}(2008)\citenamefont
  {Andreev}, \citenamefont {Labzowsky}, \citenamefont {Plunien},\ and\
  \citenamefont {Solovyev}}]{andreev08pr}%
  \BibitemOpen
  \bibfield  {author} {\bibinfo {author} {\bibfnamefont {O.~Y.}\ \bibnamefont
  {Andreev}}, \bibinfo {author} {\bibfnamefont {L.~N.}\ \bibnamefont
  {Labzowsky}}, \bibinfo {author} {\bibfnamefont {G.}~\bibnamefont {Plunien}},\
  and\ \bibinfo {author} {\bibfnamefont {D.~A.}\ \bibnamefont {Solovyev}},\
  }\bibfield  {title} {\bibinfo {title} {{QED} theory of the spectral line
  profile and its applications to atoms and ions},\ }\href
  {https://doi.org/https://doi.org/10.1016/j.physrep.2007.10.003} {\bibfield
  {journal} {\bibinfo  {journal} {Physics Reports}\ }\textbf {\bibinfo {volume}
  {455}},\ \bibinfo {pages} {135 } (\bibinfo {year} {2008})}\BibitemShut
  {NoStop}%
\bibitem [{\citenamefont {Andreev}\ \emph {et~al.}(2009)\citenamefont
  {Andreev}, \citenamefont {Labzowsky},\ and\ \citenamefont
  {Prigorovsky}}]{andreev09p042514}%
  \BibitemOpen
  \bibfield  {author} {\bibinfo {author} {\bibfnamefont {O.~Y.}\ \bibnamefont
  {Andreev}}, \bibinfo {author} {\bibfnamefont {L.~N.}\ \bibnamefont
  {Labzowsky}},\ and\ \bibinfo {author} {\bibfnamefont {A.~V.}\ \bibnamefont
  {Prigorovsky}},\ }\bibfield  {title} {\bibinfo {title} {Line-profile approach
  to the description of the electron-recombination process for the highly
  charged ions},\ }\href {https://doi.org/10.1103/PhysRevA.80.042514}
  {\bibfield  {journal} {\bibinfo  {journal} {Phys. Rev. A}\ }\textbf {\bibinfo
  {volume} {80}},\ \bibinfo {pages} {042514} (\bibinfo {year}
  {2009})}\BibitemShut {NoStop}%
\bibitem [{\citenamefont {Vasileva}\ \emph {et~al.}(2021)\citenamefont
  {Vasileva}, \citenamefont {Lyashchenko}, \citenamefont {Voitkiv},
  \citenamefont {Yu},\ and\ \citenamefont
  {Andreev}}]{Vasileva2021PhysRevA.104.052808}%
  \BibitemOpen
  \bibfield  {author} {\bibinfo {author} {\bibfnamefont {D.~M.}\ \bibnamefont
  {Vasileva}}, \bibinfo {author} {\bibfnamefont {K.~N.}\ \bibnamefont
  {Lyashchenko}}, \bibinfo {author} {\bibfnamefont {A.~B.}\ \bibnamefont
  {Voitkiv}}, \bibinfo {author} {\bibfnamefont {D.}~\bibnamefont {Yu}},\ and\
  \bibinfo {author} {\bibfnamefont {O.~Y.}\ \bibnamefont {Andreev}},\
  }\bibfield  {title} {\bibinfo {title} {Resonant elastic scattering of
  polarized electrons on h-like ions},\ }\href
  {https://doi.org/10.1103/PhysRevA.104.052808} {\bibfield  {journal} {\bibinfo
   {journal} {Phys. Rev. A}\ }\textbf {\bibinfo {volume} {104}},\ \bibinfo
  {pages} {052808} (\bibinfo {year} {2021})}\BibitemShut {NoStop}%
\bibitem [{\citenamefont {Vasileva}\ \emph {et~al.}(2024)\citenamefont
  {Vasileva}, \citenamefont {Lyashchenko}, \citenamefont {Andreev},\ and\
  \citenamefont {Yu}}]{Vasileva_2024}%
  \BibitemOpen
  \bibfield  {author} {\bibinfo {author} {\bibfnamefont {D.~M.}\ \bibnamefont
  {Vasileva}}, \bibinfo {author} {\bibfnamefont {K.~N.}\ \bibnamefont
  {Lyashchenko}}, \bibinfo {author} {\bibfnamefont {O.~Y.}\ \bibnamefont
  {Andreev}},\ and\ \bibinfo {author} {\bibfnamefont {D.}~\bibnamefont {Yu}},\
  }\bibfield  {title} {\bibinfo {title} {Inelastic resonant scattering of
  electrons on hydrogen-like ions},\ }\href
  {https://doi.org/10.1088/1402-4896/ad87b9} {\bibfield  {journal} {\bibinfo
  {journal} {Physica Scripta}\ }\textbf {\bibinfo {volume} {99}},\ \bibinfo
  {pages} {115410} (\bibinfo {year} {2024})}\BibitemShut {NoStop}%
\bibitem [{\citenamefont {Knyazeva}\ \emph {et~al.}(2022)\citenamefont
  {Knyazeva}, \citenamefont {Lyashchenko}, \citenamefont {Zhang}, \citenamefont
  {Yu},\ and\ \citenamefont {Andreev}}]{Knyazeva2022PhysRevA.106.012809}%
  \BibitemOpen
  \bibfield  {author} {\bibinfo {author} {\bibfnamefont {V.~A.}\ \bibnamefont
  {Knyazeva}}, \bibinfo {author} {\bibfnamefont {K.~N.}\ \bibnamefont
  {Lyashchenko}}, \bibinfo {author} {\bibfnamefont {M.}~\bibnamefont {Zhang}},
  \bibinfo {author} {\bibfnamefont {D.}~\bibnamefont {Yu}},\ and\ \bibinfo
  {author} {\bibfnamefont {O.~Y.}\ \bibnamefont {Andreev}},\ }\bibfield
  {title} {\bibinfo {title} {Investigation of two-photon
  $2s\ensuremath{\rightarrow}1s$ decay in one-electron and one-muon ions},\
  }\href {https://doi.org/10.1103/PhysRevA.106.012809} {\bibfield  {journal}
  {\bibinfo  {journal} {Phys. Rev. A}\ }\textbf {\bibinfo {volume} {106}},\
  \bibinfo {pages} {012809} (\bibinfo {year} {2022})}\BibitemShut {NoStop}%
\bibitem [{\citenamefont {Gorringe}\ and\ \citenamefont
  {Hertzog}(2015)}]{GORRINGE201573}%
  \BibitemOpen
  \bibfield  {author} {\bibinfo {author} {\bibfnamefont {T.}~\bibnamefont
  {Gorringe}}\ and\ \bibinfo {author} {\bibfnamefont {D.}~\bibnamefont
  {Hertzog}},\ }\bibfield  {title} {\bibinfo {title} {Precision muon physics},\
  }\href {https://doi.org/https://doi.org/10.1016/j.ppnp.2015.06.001}
  {\bibfield  {journal} {\bibinfo  {journal} {Progress in Particle and Nuclear
  Physics}\ }\textbf {\bibinfo {volume} {84}},\ \bibinfo {pages} {73} (\bibinfo
  {year} {2015})}\BibitemShut {NoStop}%
\bibitem [{\citenamefont {Cline}\ \emph {et~al.}(2022)\citenamefont {Cline},
  \citenamefont {Lin}, \citenamefont {Roy}, \citenamefont {Reimer},
  \citenamefont {Mesick}, \citenamefont {Akmal}, \citenamefont {Alie},
  \citenamefont {Atac}, \citenamefont {Atencio}, \citenamefont {Ayerbe~Gayoso},
  \citenamefont {Benmouna}, \citenamefont {Benmokhtar}, \citenamefont
  {Bernauer}, \citenamefont {Briscoe}, \citenamefont {Campbell}, \citenamefont
  {Cohen}, \citenamefont {Cohen}, \citenamefont {Collicott}, \citenamefont
  {Deiters}, \citenamefont {Dogra}, \citenamefont {Downie}, \citenamefont
  {Fernando}, \citenamefont {Flannery}, \citenamefont {Gautam}, \citenamefont
  {Ghosal}, \citenamefont {Gilman}, \citenamefont {Golossanov}, \citenamefont
  {Halter}, \citenamefont {Hirschman}, \citenamefont {Ilieva}, \citenamefont
  {Kim}, \citenamefont {Kohl}, \citenamefont {Krusche}, \citenamefont
  {Lavrukhin}, \citenamefont {Li}, \citenamefont {Liang-Gilman}, \citenamefont
  {Liyanage}, \citenamefont {Lorenzon}, \citenamefont {Mohanmurthy},
  \citenamefont {Mokal}, \citenamefont {Moran}, \citenamefont {Nazeer},
  \citenamefont {Or}, \citenamefont {Patel}, \citenamefont {Piasetzky},
  \citenamefont {Rauber}, \citenamefont {Raymond}, \citenamefont {Reggiani},
  \citenamefont {Reid}, \citenamefont {Ron}, \citenamefont {Rooney},
  \citenamefont {Rostomyan}, \citenamefont {Schwarz}, \citenamefont {Sneath},
  \citenamefont {Solazzo}, \citenamefont {Sparveris}, \citenamefont
  {Steinberg}, \citenamefont {Strauch}, \citenamefont {Sulkosky},\ and\
  \citenamefont {Wuerfel}}]{Cline2022PhysRevC.105.055201}%
  \BibitemOpen
  \bibfield  {author} {\bibinfo {author} {\bibfnamefont {E.}~\bibnamefont
  {Cline}}, \bibinfo {author} {\bibfnamefont {W.}~\bibnamefont {Lin}}, \bibinfo
  {author} {\bibfnamefont {P.}~\bibnamefont {Roy}}, \bibinfo {author}
  {\bibfnamefont {P.~E.}\ \bibnamefont {Reimer}}, \bibinfo {author}
  {\bibfnamefont {K.~E.}\ \bibnamefont {Mesick}}, \bibinfo {author}
  {\bibfnamefont {A.}~\bibnamefont {Akmal}}, \bibinfo {author} {\bibfnamefont
  {A.}~\bibnamefont {Alie}}, \bibinfo {author} {\bibfnamefont {H.}~\bibnamefont
  {Atac}}, \bibinfo {author} {\bibfnamefont {A.}~\bibnamefont {Atencio}},
  \bibinfo {author} {\bibfnamefont {C.}~\bibnamefont {Ayerbe~Gayoso}}, \bibinfo
  {author} {\bibfnamefont {N.}~\bibnamefont {Benmouna}}, \bibinfo {author}
  {\bibfnamefont {F.}~\bibnamefont {Benmokhtar}}, \bibinfo {author}
  {\bibfnamefont {J.~C.}\ \bibnamefont {Bernauer}}, \bibinfo {author}
  {\bibfnamefont {W.~J.}\ \bibnamefont {Briscoe}}, \bibinfo {author}
  {\bibfnamefont {J.}~\bibnamefont {Campbell}}, \bibinfo {author}
  {\bibfnamefont {D.}~\bibnamefont {Cohen}}, \bibinfo {author} {\bibfnamefont
  {E.~O.}\ \bibnamefont {Cohen}}, \bibinfo {author} {\bibfnamefont
  {C.}~\bibnamefont {Collicott}}, \bibinfo {author} {\bibfnamefont
  {K.}~\bibnamefont {Deiters}}, \bibinfo {author} {\bibfnamefont
  {S.}~\bibnamefont {Dogra}}, \bibinfo {author} {\bibfnamefont
  {E.}~\bibnamefont {Downie}}, \bibinfo {author} {\bibfnamefont {I.~P.}\
  \bibnamefont {Fernando}}, \bibinfo {author} {\bibfnamefont {A.}~\bibnamefont
  {Flannery}}, \bibinfo {author} {\bibfnamefont {T.}~\bibnamefont {Gautam}},
  \bibinfo {author} {\bibfnamefont {D.}~\bibnamefont {Ghosal}}, \bibinfo
  {author} {\bibfnamefont {R.}~\bibnamefont {Gilman}}, \bibinfo {author}
  {\bibfnamefont {A.}~\bibnamefont {Golossanov}}, \bibinfo {author}
  {\bibfnamefont {B.~F.}\ \bibnamefont {Halter}}, \bibinfo {author}
  {\bibfnamefont {J.}~\bibnamefont {Hirschman}}, \bibinfo {author}
  {\bibfnamefont {Y.}~\bibnamefont {Ilieva}}, \bibinfo {author} {\bibfnamefont
  {M.}~\bibnamefont {Kim}}, \bibinfo {author} {\bibfnamefont {M.}~\bibnamefont
  {Kohl}}, \bibinfo {author} {\bibfnamefont {B.}~\bibnamefont {Krusche}},
  \bibinfo {author} {\bibfnamefont {I.}~\bibnamefont {Lavrukhin}}, \bibinfo
  {author} {\bibfnamefont {L.}~\bibnamefont {Li}}, \bibinfo {author}
  {\bibfnamefont {B.}~\bibnamefont {Liang-Gilman}}, \bibinfo {author}
  {\bibfnamefont {A.}~\bibnamefont {Liyanage}}, \bibinfo {author}
  {\bibfnamefont {W.}~\bibnamefont {Lorenzon}}, \bibinfo {author}
  {\bibfnamefont {P.}~\bibnamefont {Mohanmurthy}}, \bibinfo {author}
  {\bibfnamefont {R.}~\bibnamefont {Mokal}}, \bibinfo {author} {\bibfnamefont
  {P.}~\bibnamefont {Moran}}, \bibinfo {author} {\bibfnamefont {S.~J.}\
  \bibnamefont {Nazeer}}, \bibinfo {author} {\bibfnamefont {P.}~\bibnamefont
  {Or}}, \bibinfo {author} {\bibfnamefont {T.}~\bibnamefont {Patel}}, \bibinfo
  {author} {\bibfnamefont {E.}~\bibnamefont {Piasetzky}}, \bibinfo {author}
  {\bibfnamefont {T.}~\bibnamefont {Rauber}}, \bibinfo {author} {\bibfnamefont
  {R.~S.}\ \bibnamefont {Raymond}}, \bibinfo {author} {\bibfnamefont
  {D.}~\bibnamefont {Reggiani}}, \bibinfo {author} {\bibfnamefont
  {H.}~\bibnamefont {Reid}}, \bibinfo {author} {\bibfnamefont {G.}~\bibnamefont
  {Ron}}, \bibinfo {author} {\bibfnamefont {E.}~\bibnamefont {Rooney}},
  \bibinfo {author} {\bibfnamefont {T.}~\bibnamefont {Rostomyan}}, \bibinfo
  {author} {\bibfnamefont {M.}~\bibnamefont {Schwarz}}, \bibinfo {author}
  {\bibfnamefont {A.}~\bibnamefont {Sneath}}, \bibinfo {author} {\bibfnamefont
  {P.~G.}\ \bibnamefont {Solazzo}}, \bibinfo {author} {\bibfnamefont
  {N.}~\bibnamefont {Sparveris}}, \bibinfo {author} {\bibfnamefont
  {N.}~\bibnamefont {Steinberg}}, \bibinfo {author} {\bibfnamefont
  {S.}~\bibnamefont {Strauch}}, \bibinfo {author} {\bibfnamefont
  {V.}~\bibnamefont {Sulkosky}},\ and\ \bibinfo {author} {\bibfnamefont
  {N.}~\bibnamefont {Wuerfel}},\ }\bibfield  {title} {\bibinfo {title}
  {Characterization of muon and electron beams in the paul scherrer institute
  pim1 channel for the muse experiment},\ }\href
  {https://doi.org/10.1103/PhysRevC.105.055201} {\bibfield  {journal} {\bibinfo
   {journal} {Phys. Rev. C}\ }\textbf {\bibinfo {volume} {105}},\ \bibinfo
  {pages} {055201} (\bibinfo {year} {2022})}\BibitemShut {NoStop}%
\bibitem [{\citenamefont {Bao}\ \emph {et~al.}(2023)\citenamefont {Bao},
  \citenamefont {Chen}, \citenamefont {Chen}, \citenamefont {Cheng},
  \citenamefont {Deng}, \citenamefont {Fan}, \citenamefont {Guo}, \citenamefont
  {He}, \citenamefont {Hu}, \citenamefont {Li}, \citenamefont {Li},
  \citenamefont {Liang}, \citenamefont {Liu}, \citenamefont {Lv}, \citenamefont
  {Pan}, \citenamefont {Tan}, \citenamefont {Vassilopoulos}, \citenamefont
  {Wu}, \citenamefont {Yang},\ and\ \citenamefont {Zhang}}]{Bao_2023}%
  \BibitemOpen
  \bibfield  {author} {\bibinfo {author} {\bibfnamefont {Y.}~\bibnamefont
  {Bao}}, \bibinfo {author} {\bibfnamefont {J.}~\bibnamefont {Chen}}, \bibinfo
  {author} {\bibfnamefont {C.}~\bibnamefont {Chen}}, \bibinfo {author}
  {\bibfnamefont {H.}~\bibnamefont {Cheng}}, \bibinfo {author} {\bibfnamefont
  {C.}~\bibnamefont {Deng}}, \bibinfo {author} {\bibfnamefont {R.}~\bibnamefont
  {Fan}}, \bibinfo {author} {\bibfnamefont {Y.}~\bibnamefont {Guo}}, \bibinfo
  {author} {\bibfnamefont {N.}~\bibnamefont {He}}, \bibinfo {author}
  {\bibfnamefont {H.}~\bibnamefont {Hu}}, \bibinfo {author} {\bibfnamefont
  {Q.}~\bibnamefont {Li}}, \bibinfo {author} {\bibfnamefont {Y.}~\bibnamefont
  {Li}}, \bibinfo {author} {\bibfnamefont {H.}~\bibnamefont {Liang}}, \bibinfo
  {author} {\bibfnamefont {L.}~\bibnamefont {Liu}}, \bibinfo {author}
  {\bibfnamefont {Y.}~\bibnamefont {Lv}}, \bibinfo {author} {\bibfnamefont
  {Z.}~\bibnamefont {Pan}}, \bibinfo {author} {\bibfnamefont {Z.}~\bibnamefont
  {Tan}}, \bibinfo {author} {\bibfnamefont {N.}~\bibnamefont {Vassilopoulos}},
  \bibinfo {author} {\bibfnamefont {Y.}~\bibnamefont {Wu}}, \bibinfo {author}
  {\bibfnamefont {T.}~\bibnamefont {Yang}},\ and\ \bibinfo {author}
  {\bibfnamefont {G.}~\bibnamefont {Zhang}},\ }\bibfield  {title} {\bibinfo
  {title} {Progress report on muon source project at csns},\ }\href
  {https://doi.org/10.1088/1742-6596/2462/1/012034} {\bibfield  {journal}
  {\bibinfo  {journal} {Journal of Physics: Conference Series}\ }\textbf
  {\bibinfo {volume} {2462}},\ \bibinfo {pages} {012034} (\bibinfo {year}
  {2023})}\BibitemShut {NoStop}%
\bibitem [{\citenamefont {Cai}\ \emph {et~al.}(2024)\citenamefont {Cai},
  \citenamefont {He}, \citenamefont {Liu}, \citenamefont {Jia}, \citenamefont
  {Qin}, \citenamefont {Wang}, \citenamefont {Wang}, \citenamefont {Zhao},
  \citenamefont {Pu}, \citenamefont {Niu}, \citenamefont {Chen}, \citenamefont
  {Sun}, \citenamefont {Zhao},\ and\ \citenamefont
  {Zhan}}]{Cai2024PhysRevAccelBeams.27.023403}%
  \BibitemOpen
  \bibfield  {author} {\bibinfo {author} {\bibfnamefont {H.-J.}\ \bibnamefont
  {Cai}}, \bibinfo {author} {\bibfnamefont {Y.}~\bibnamefont {He}}, \bibinfo
  {author} {\bibfnamefont {S.}~\bibnamefont {Liu}}, \bibinfo {author}
  {\bibfnamefont {H.}~\bibnamefont {Jia}}, \bibinfo {author} {\bibfnamefont
  {Y.}~\bibnamefont {Qin}}, \bibinfo {author} {\bibfnamefont {Z.}~\bibnamefont
  {Wang}}, \bibinfo {author} {\bibfnamefont {F.}~\bibnamefont {Wang}}, \bibinfo
  {author} {\bibfnamefont {L.}~\bibnamefont {Zhao}}, \bibinfo {author}
  {\bibfnamefont {N.}~\bibnamefont {Pu}}, \bibinfo {author} {\bibfnamefont
  {J.}~\bibnamefont {Niu}}, \bibinfo {author} {\bibfnamefont {L.}~\bibnamefont
  {Chen}}, \bibinfo {author} {\bibfnamefont {Z.}~\bibnamefont {Sun}}, \bibinfo
  {author} {\bibfnamefont {H.}~\bibnamefont {Zhao}},\ and\ \bibinfo {author}
  {\bibfnamefont {W.}~\bibnamefont {Zhan}},\ }\bibfield  {title} {\bibinfo
  {title} {Towards a high-intensity muon source},\ }\href
  {https://doi.org/10.1103/PhysRevAccelBeams.27.023403} {\bibfield  {journal}
  {\bibinfo  {journal} {Phys. Rev. Accel. Beams}\ }\textbf {\bibinfo {volume}
  {27}},\ \bibinfo {pages} {023403} (\bibinfo {year} {2024})}\BibitemShut
  {NoStop}%
\bibitem [{\citenamefont {{Adamczak, A.}}\ \emph {et~al.}(2018)\citenamefont
  {{Adamczak, A.}}, \citenamefont {{Antognini, A.}}, \citenamefont {{Berger,
  N.}}, \citenamefont {{Cocolios, T.E.}}, \citenamefont {{Dressler, R.}},
  \citenamefont {{Eggenberger, A.}}, \citenamefont {{Eichler, R.}},
  \citenamefont {{Indelicato, P.}}, \citenamefont {{Jungmann, K.}},
  \citenamefont {{Kirch, K.}}, \citenamefont {{Knecht, A.}}, \citenamefont
  {{Papa, A.}}, \citenamefont {{Pohl, R.}}, \citenamefont {{Pospelov, M.}},
  \citenamefont {{Rapisarda, E.}}, \citenamefont {{Reiter, P.}}, \citenamefont
  {{Ritjoho, N.}}, \citenamefont {{Roccia, S.}}, \citenamefont {{Severijns,
  N.}}, \citenamefont {{Skawran, A.}}, \citenamefont {{Wauters, F.}},\ and\
  \citenamefont {{Willmann, L.}}}]{Adamczak2018refId0}%
  \BibitemOpen
  \bibfield  {author} {\bibinfo {author} {\bibnamefont {{Adamczak, A.}}},
  \bibinfo {author} {\bibnamefont {{Antognini, A.}}}, \bibinfo {author}
  {\bibnamefont {{Berger, N.}}}, \bibinfo {author} {\bibnamefont {{Cocolios,
  T.E.}}}, \bibinfo {author} {\bibnamefont {{Dressler, R.}}}, \bibinfo {author}
  {\bibnamefont {{Eggenberger, A.}}}, \bibinfo {author} {\bibnamefont
  {{Eichler, R.}}}, \bibinfo {author} {\bibnamefont {{Indelicato, P.}}},
  \bibinfo {author} {\bibnamefont {{Jungmann, K.}}}, \bibinfo {author}
  {\bibnamefont {{Kirch, K.}}}, \bibinfo {author} {\bibnamefont {{Knecht,
  A.}}}, \bibinfo {author} {\bibnamefont {{Papa, A.}}}, \bibinfo {author}
  {\bibnamefont {{Pohl, R.}}}, \bibinfo {author} {\bibnamefont {{Pospelov,
  M.}}}, \bibinfo {author} {\bibnamefont {{Rapisarda, E.}}}, \bibinfo {author}
  {\bibnamefont {{Reiter, P.}}}, \bibinfo {author} {\bibnamefont {{Ritjoho,
  N.}}}, \bibinfo {author} {\bibnamefont {{Roccia, S.}}}, \bibinfo {author}
  {\bibnamefont {{Severijns, N.}}}, \bibinfo {author} {\bibnamefont {{Skawran,
  A.}}}, \bibinfo {author} {\bibnamefont {{Wauters, F.}}},\ and\ \bibinfo
  {author} {\bibnamefont {{Willmann, L.}}},\ }\bibfield  {title} {\bibinfo
  {title} {Nuclear structure with radioactive muonic atoms},\ }\href
  {https://doi.org/10.1051/epjconf/201819304014} {\bibfield  {journal}
  {\bibinfo  {journal} {EPJ Web Conf.}\ }\textbf {\bibinfo {volume} {193}},\
  \bibinfo {pages} {04014} (\bibinfo {year} {2018})}\BibitemShut {NoStop}%
\bibitem [{\citenamefont {Antognini}\ \emph {et~al.}(2020)\citenamefont
  {Antognini}, \citenamefont {Berger}, \citenamefont {Cocolios}, \citenamefont
  {Dressler}, \citenamefont {Eichler}, \citenamefont {Eggenberger},
  \citenamefont {Indelicato}, \citenamefont {Jungmann}, \citenamefont {Keitel},
  \citenamefont {Kirch}, \citenamefont {Knecht}, \citenamefont {Michel},
  \citenamefont {Nuber}, \citenamefont {Oreshkina}, \citenamefont {Ouf},
  \citenamefont {Papa}, \citenamefont {Pohl}, \citenamefont {Pospelov},
  \citenamefont {Rapisarda}, \citenamefont {Ritjoho}, \citenamefont {Roccia},
  \citenamefont {Severijns}, \citenamefont {Skawran}, \citenamefont {Vogiatzi},
  \citenamefont {Wauters},\ and\ \citenamefont
  {Willmann}}]{Antognini2020PhysRevC.101.054313}%
  \BibitemOpen
  \bibfield  {author} {\bibinfo {author} {\bibfnamefont {A.}~\bibnamefont
  {Antognini}}, \bibinfo {author} {\bibfnamefont {N.}~\bibnamefont {Berger}},
  \bibinfo {author} {\bibfnamefont {T.~E.}\ \bibnamefont {Cocolios}}, \bibinfo
  {author} {\bibfnamefont {R.}~\bibnamefont {Dressler}}, \bibinfo {author}
  {\bibfnamefont {R.}~\bibnamefont {Eichler}}, \bibinfo {author} {\bibfnamefont
  {A.}~\bibnamefont {Eggenberger}}, \bibinfo {author} {\bibfnamefont
  {P.}~\bibnamefont {Indelicato}}, \bibinfo {author} {\bibfnamefont
  {K.}~\bibnamefont {Jungmann}}, \bibinfo {author} {\bibfnamefont {C.~H.}\
  \bibnamefont {Keitel}}, \bibinfo {author} {\bibfnamefont {K.}~\bibnamefont
  {Kirch}}, \bibinfo {author} {\bibfnamefont {A.}~\bibnamefont {Knecht}},
  \bibinfo {author} {\bibfnamefont {N.}~\bibnamefont {Michel}}, \bibinfo
  {author} {\bibfnamefont {J.}~\bibnamefont {Nuber}}, \bibinfo {author}
  {\bibfnamefont {N.~S.}\ \bibnamefont {Oreshkina}}, \bibinfo {author}
  {\bibfnamefont {A.}~\bibnamefont {Ouf}}, \bibinfo {author} {\bibfnamefont
  {A.}~\bibnamefont {Papa}}, \bibinfo {author} {\bibfnamefont {R.}~\bibnamefont
  {Pohl}}, \bibinfo {author} {\bibfnamefont {M.}~\bibnamefont {Pospelov}},
  \bibinfo {author} {\bibfnamefont {E.}~\bibnamefont {Rapisarda}}, \bibinfo
  {author} {\bibfnamefont {N.}~\bibnamefont {Ritjoho}}, \bibinfo {author}
  {\bibfnamefont {S.}~\bibnamefont {Roccia}}, \bibinfo {author} {\bibfnamefont
  {N.}~\bibnamefont {Severijns}}, \bibinfo {author} {\bibfnamefont
  {A.}~\bibnamefont {Skawran}}, \bibinfo {author} {\bibfnamefont {S.~M.}\
  \bibnamefont {Vogiatzi}}, \bibinfo {author} {\bibfnamefont {F.}~\bibnamefont
  {Wauters}},\ and\ \bibinfo {author} {\bibfnamefont {L.}~\bibnamefont
  {Willmann}},\ }\bibfield  {title} {\bibinfo {title} {Measurement of the
  quadrupole moment of $^{185}\mathrm{Re}$ and $^{187}\mathrm{Re}$ from the
  hyperfine structure of muonic x rays},\ }\href
  {https://doi.org/10.1103/PhysRevC.101.054313} {\bibfield  {journal} {\bibinfo
   {journal} {Phys. Rev. C}\ }\textbf {\bibinfo {volume} {101}},\ \bibinfo
  {pages} {054313} (\bibinfo {year} {2020})}\BibitemShut {NoStop}%
\bibitem [{\citenamefont {Paul}\ \emph {et~al.}(2021)\citenamefont {Paul},
  \citenamefont {Bian}, \citenamefont {Azuma}, \citenamefont {Okada},\ and\
  \citenamefont {Indelicato}}]{Paul2021PhysRevLett.126.173001}%
  \BibitemOpen
  \bibfield  {author} {\bibinfo {author} {\bibfnamefont {N.}~\bibnamefont
  {Paul}}, \bibinfo {author} {\bibfnamefont {G.}~\bibnamefont {Bian}}, \bibinfo
  {author} {\bibfnamefont {T.}~\bibnamefont {Azuma}}, \bibinfo {author}
  {\bibfnamefont {S.}~\bibnamefont {Okada}},\ and\ \bibinfo {author}
  {\bibfnamefont {P.}~\bibnamefont {Indelicato}},\ }\bibfield  {title}
  {\bibinfo {title} {Testing quantum electrodynamics with exotic atoms},\
  }\href {https://doi.org/10.1103/PhysRevLett.126.173001} {\bibfield  {journal}
  {\bibinfo  {journal} {Phys. Rev. Lett.}\ }\textbf {\bibinfo {volume} {126}},\
  \bibinfo {pages} {173001} (\bibinfo {year} {2021})}\BibitemShut {NoStop}%
\bibitem [{\citenamefont {Andreev}\ \emph {et~al.}(2025)\citenamefont
  {Andreev}, \citenamefont {Yu}, \citenamefont {Lyashchenko},\ and\
  \citenamefont {Vasileva}}]{andreev2025cpl}%
  \BibitemOpen
  \bibfield  {author} {\bibinfo {author} {\bibfnamefont {O.~Y.}\ \bibnamefont
  {Andreev}}, \bibinfo {author} {\bibfnamefont {D.}~\bibnamefont {Yu}},
  \bibinfo {author} {\bibfnamefont {K.~N.}\ \bibnamefont {Lyashchenko}},\ and\
  \bibinfo {author} {\bibfnamefont {D.~M.}\ \bibnamefont {Vasileva}},\
  }\bibfield  {title} {\bibinfo {title} {Importance of the breit interaction
  for electron–positron pair production in bound–bound muon transitions},\
  }\href {https://doi.org/10.1088/0256-307X/42/9/090301} {\bibfield  {journal}
  {\bibinfo  {journal} {Chin. Phys. Lett.}\ }\textbf {\bibinfo {volume} {42}},\
  \bibinfo {pages} {090301} (\bibinfo {year} {2025})}\BibitemShut {NoStop}%
\bibitem [{\citenamefont {Andreev}\ \emph {et~al.}(2026)\citenamefont
  {Andreev}, \citenamefont {Yu}, \citenamefont {Lyashchenko},\ and\
  \citenamefont {Vasileva}}]{andreev2026muonenergy}%
  \BibitemOpen
  \bibfield  {author} {\bibinfo {author} {\bibfnamefont {O.~Y.}\ \bibnamefont
  {Andreev}}, \bibinfo {author} {\bibfnamefont {D.}~\bibnamefont {Yu}},
  \bibinfo {author} {\bibfnamefont {K.~N.}\ \bibnamefont {Lyashchenko}},\ and\
  \bibinfo {author} {\bibfnamefont {D.~M.}\ \bibnamefont {Vasileva}},\
  }\bibfield  {title} {\bibinfo {title} {Energy spectra of electron-positron
  pairs produced in bound-bound muon transitions},\ }\href
  {https://doi.org/10.1103/m9l8-6g96} {\bibfield  {journal} {\bibinfo
  {journal} {Phys. Rev. A}\ }\textbf {\bibinfo {volume} {113}},\ \bibinfo
  {pages} {042804} (\bibinfo {year} {2026})}\BibitemShut {NoStop}%
\bibitem [{\citenamefont {Hofstadter}(1956)}]{Hofstadter1956RevModPhys.28.214}%
  \BibitemOpen
  \bibfield  {author} {\bibinfo {author} {\bibfnamefont {R.}~\bibnamefont
  {Hofstadter}},\ }\bibfield  {title} {\bibinfo {title} {Electron scattering
  and nuclear structure},\ }\href {https://doi.org/10.1103/RevModPhys.28.214}
  {\bibfield  {journal} {\bibinfo  {journal} {Rev. Mod. Phys.}\ }\textbf
  {\bibinfo {volume} {28}},\ \bibinfo {pages} {214} (\bibinfo {year}
  {1956})}\BibitemShut {NoStop}%
\bibitem [{\citenamefont {Barrett}(1974)}]{Barrett_1974}%
  \BibitemOpen
  \bibfield  {author} {\bibinfo {author} {\bibfnamefont {R.~C.}\ \bibnamefont
  {Barrett}},\ }\bibfield  {title} {\bibinfo {title} {Nuclear charge
  distributions},\ }\href {https://doi.org/10.1088/0034-4885/37/1/001}
  {\bibfield  {journal} {\bibinfo  {journal} {Reports on Progress in Physics}\
  }\textbf {\bibinfo {volume} {37}},\ \bibinfo {pages} {1} (\bibinfo {year}
  {1974})}\BibitemShut {NoStop}%
\bibitem [{\citenamefont {{De Vries}}\ \emph {et~al.}(1987)\citenamefont {{De
  Vries}}, \citenamefont {{De Jager}},\ and\ \citenamefont {{De
  Vries}}}]{DEVRIES1987495}%
  \BibitemOpen
  \bibfield  {author} {\bibinfo {author} {\bibfnamefont {H.}~\bibnamefont {{De
  Vries}}}, \bibinfo {author} {\bibfnamefont {C.}~\bibnamefont {{De Jager}}},\
  and\ \bibinfo {author} {\bibfnamefont {C.}~\bibnamefont {{De Vries}}},\
  }\bibfield  {title} {\bibinfo {title} {Nuclear charge-density-distribution
  parameters from elastic electron scattering},\ }\href
  {https://doi.org/https://doi.org/10.1016/0092-640X(87)90013-1} {\bibfield
  {journal} {\bibinfo  {journal} {Atomic Data and Nuclear Data Tables}\
  }\textbf {\bibinfo {volume} {36}},\ \bibinfo {pages} {495} (\bibinfo {year}
  {1987})}\BibitemShut {NoStop}%
\bibitem [{\citenamefont {Skripnikov}\ \emph {et~al.}(2018)\citenamefont
  {Skripnikov}, \citenamefont {Schmidt}, \citenamefont {Ullmann}, \citenamefont
  {Geppert}, \citenamefont {Kraus}, \citenamefont {Kresse}, \citenamefont
  {N\"ortersh\"auser}, \citenamefont {Privalov}, \citenamefont {Scheibe},
  \citenamefont {Shabaev}, \citenamefont {Vogel},\ and\ \citenamefont
  {Volotka}}]{Skripnikov2018PhysRevLett.120.093001}%
  \BibitemOpen
  \bibfield  {author} {\bibinfo {author} {\bibfnamefont {L.~V.}\ \bibnamefont
  {Skripnikov}}, \bibinfo {author} {\bibfnamefont {S.}~\bibnamefont {Schmidt}},
  \bibinfo {author} {\bibfnamefont {J.}~\bibnamefont {Ullmann}}, \bibinfo
  {author} {\bibfnamefont {C.}~\bibnamefont {Geppert}}, \bibinfo {author}
  {\bibfnamefont {F.}~\bibnamefont {Kraus}}, \bibinfo {author} {\bibfnamefont
  {B.}~\bibnamefont {Kresse}}, \bibinfo {author} {\bibfnamefont
  {W.}~\bibnamefont {N\"ortersh\"auser}}, \bibinfo {author} {\bibfnamefont
  {A.~F.}\ \bibnamefont {Privalov}}, \bibinfo {author} {\bibfnamefont
  {B.}~\bibnamefont {Scheibe}}, \bibinfo {author} {\bibfnamefont {V.~M.}\
  \bibnamefont {Shabaev}}, \bibinfo {author} {\bibfnamefont {M.}~\bibnamefont
  {Vogel}},\ and\ \bibinfo {author} {\bibfnamefont {A.~V.}\ \bibnamefont
  {Volotka}},\ }\bibfield  {title} {\bibinfo {title} {New nuclear magnetic
  moment of $^{209}\mathrm{Bi}$: Resolving the bismuth hyperfine puzzle},\
  }\href {https://doi.org/10.1103/PhysRevLett.120.093001} {\bibfield  {journal}
  {\bibinfo  {journal} {Phys. Rev. Lett.}\ }\textbf {\bibinfo {volume} {120}},\
  \bibinfo {pages} {093001} (\bibinfo {year} {2018})}\BibitemShut {NoStop}%
\bibitem [{\citenamefont {Stone}(2025)}]{stone_2025_16kpj-2k407}%
  \BibitemOpen
  \bibfield  {author} {\bibinfo {author} {\bibfnamefont {N.}~\bibnamefont
  {Stone}},\ }\href {https://doi.org/10.61092/iaea.qs1q-27sa} {\bibinfo {title}
  {Table of recommended nuclear magnetic dipole moments: Part i, long-lived
  states}} (\bibinfo {year} {2025})\BibitemShut {NoStop}%
\bibitem [{\citenamefont {Gustafsson}\ \emph {et~al.}(2025)\citenamefont
  {Gustafsson}, \citenamefont {Rodr\'{\i}guez}, \citenamefont {Garcia~Ruiz},
  \citenamefont {Miyagi}, \citenamefont {Bai}, \citenamefont {Balabanski},
  \citenamefont {Binnersley}, \citenamefont {Bissell}, \citenamefont {Blaum},
  \citenamefont {Cheal}, \citenamefont {Cocolios}, \citenamefont
  {Farooq-Smith}, \citenamefont {Flanagan}, \citenamefont {Franchoo},
  \citenamefont {Galindo-Uribarri}, \citenamefont {Georgiev}, \citenamefont
  {Gins}, \citenamefont {Gorges}, \citenamefont {de~Groote}, \citenamefont
  {Heylen}, \citenamefont {Holt}, \citenamefont {Kanellakopoulos},
  \citenamefont {Karthein}, \citenamefont {Kaufmann}, \citenamefont
  {Koszor\'us}, \citenamefont {K\"onig}, \citenamefont {Lagaki}, \citenamefont
  {Lechner}, \citenamefont {Maass}, \citenamefont {Malbrunot-Ettenauer},
  \citenamefont {Nazarewicz}, \citenamefont {Neugart}, \citenamefont {Neyens},
  \citenamefont {N\"ortersh\"auser}, \citenamefont {Otsuka}, \citenamefont
  {Reinhard}, \citenamefont {Rondelez}, \citenamefont {Romero-Romero},
  \citenamefont {Ricketts}, \citenamefont {Sailer}, \citenamefont {S\'anchez},
  \citenamefont {Schmidt}, \citenamefont {Schwenk}, \citenamefont {Stroberg},
  \citenamefont {Shimizu}, \citenamefont {Tsunoda}, \citenamefont {Vernon},
  \citenamefont {Wehner}, \citenamefont {Wilkins}, \citenamefont {Wraith},
  \citenamefont {Xie}, \citenamefont {Xu}, \citenamefont {Yang},\ and\
  \citenamefont {Yordanov}}]{Gustafsson2025wbdx-k3cd}%
  \BibitemOpen
  \bibfield  {author} {\bibinfo {author} {\bibfnamefont {F.~P.}\ \bibnamefont
  {Gustafsson}}, \bibinfo {author} {\bibfnamefont {L.~V.}\ \bibnamefont
  {Rodr\'{\i}guez}}, \bibinfo {author} {\bibfnamefont {R.~F.}\ \bibnamefont
  {Garcia~Ruiz}}, \bibinfo {author} {\bibfnamefont {T.}~\bibnamefont {Miyagi}},
  \bibinfo {author} {\bibfnamefont {S.~W.}\ \bibnamefont {Bai}}, \bibinfo
  {author} {\bibfnamefont {D.~L.}\ \bibnamefont {Balabanski}}, \bibinfo
  {author} {\bibfnamefont {C.~L.}\ \bibnamefont {Binnersley}}, \bibinfo
  {author} {\bibfnamefont {M.~L.}\ \bibnamefont {Bissell}}, \bibinfo {author}
  {\bibfnamefont {K.}~\bibnamefont {Blaum}}, \bibinfo {author} {\bibfnamefont
  {B.}~\bibnamefont {Cheal}}, \bibinfo {author} {\bibfnamefont {T.~E.}\
  \bibnamefont {Cocolios}}, \bibinfo {author} {\bibfnamefont {G.~J.}\
  \bibnamefont {Farooq-Smith}}, \bibinfo {author} {\bibfnamefont {K.~T.}\
  \bibnamefont {Flanagan}}, \bibinfo {author} {\bibfnamefont {S.}~\bibnamefont
  {Franchoo}}, \bibinfo {author} {\bibfnamefont {A.}~\bibnamefont
  {Galindo-Uribarri}}, \bibinfo {author} {\bibfnamefont {G.}~\bibnamefont
  {Georgiev}}, \bibinfo {author} {\bibfnamefont {W.}~\bibnamefont {Gins}},
  \bibinfo {author} {\bibfnamefont {C.}~\bibnamefont {Gorges}}, \bibinfo
  {author} {\bibfnamefont {R.~P.}\ \bibnamefont {de~Groote}}, \bibinfo {author}
  {\bibfnamefont {H.}~\bibnamefont {Heylen}}, \bibinfo {author} {\bibfnamefont
  {J.~D.}\ \bibnamefont {Holt}}, \bibinfo {author} {\bibfnamefont
  {A.}~\bibnamefont {Kanellakopoulos}}, \bibinfo {author} {\bibfnamefont
  {J.}~\bibnamefont {Karthein}}, \bibinfo {author} {\bibfnamefont
  {S.}~\bibnamefont {Kaufmann}}, \bibinfo {author} {\bibfnamefont
  {A.}~\bibnamefont {Koszor\'us}}, \bibinfo {author} {\bibfnamefont
  {K.}~\bibnamefont {K\"onig}}, \bibinfo {author} {\bibfnamefont
  {V.}~\bibnamefont {Lagaki}}, \bibinfo {author} {\bibfnamefont
  {S.}~\bibnamefont {Lechner}}, \bibinfo {author} {\bibfnamefont
  {B.}~\bibnamefont {Maass}}, \bibinfo {author} {\bibfnamefont
  {S.}~\bibnamefont {Malbrunot-Ettenauer}}, \bibinfo {author} {\bibfnamefont
  {W.}~\bibnamefont {Nazarewicz}}, \bibinfo {author} {\bibfnamefont
  {R.}~\bibnamefont {Neugart}}, \bibinfo {author} {\bibfnamefont
  {G.}~\bibnamefont {Neyens}}, \bibinfo {author} {\bibfnamefont
  {W.}~\bibnamefont {N\"ortersh\"auser}}, \bibinfo {author} {\bibfnamefont
  {T.}~\bibnamefont {Otsuka}}, \bibinfo {author} {\bibfnamefont {P.-G.}\
  \bibnamefont {Reinhard}}, \bibinfo {author} {\bibfnamefont {N.}~\bibnamefont
  {Rondelez}}, \bibinfo {author} {\bibfnamefont {E.}~\bibnamefont
  {Romero-Romero}}, \bibinfo {author} {\bibfnamefont {C.~M.}\ \bibnamefont
  {Ricketts}}, \bibinfo {author} {\bibfnamefont {S.}~\bibnamefont {Sailer}},
  \bibinfo {author} {\bibfnamefont {R.}~\bibnamefont {S\'anchez}}, \bibinfo
  {author} {\bibfnamefont {S.}~\bibnamefont {Schmidt}}, \bibinfo {author}
  {\bibfnamefont {A.}~\bibnamefont {Schwenk}}, \bibinfo {author} {\bibfnamefont
  {S.~R.}\ \bibnamefont {Stroberg}}, \bibinfo {author} {\bibfnamefont
  {N.}~\bibnamefont {Shimizu}}, \bibinfo {author} {\bibfnamefont
  {Y.}~\bibnamefont {Tsunoda}}, \bibinfo {author} {\bibfnamefont {A.~R.}\
  \bibnamefont {Vernon}}, \bibinfo {author} {\bibfnamefont {L.}~\bibnamefont
  {Wehner}}, \bibinfo {author} {\bibfnamefont {S.~G.}\ \bibnamefont {Wilkins}},
  \bibinfo {author} {\bibfnamefont {C.}~\bibnamefont {Wraith}}, \bibinfo
  {author} {\bibfnamefont {L.}~\bibnamefont {Xie}}, \bibinfo {author}
  {\bibfnamefont {Z.~Y.}\ \bibnamefont {Xu}}, \bibinfo {author} {\bibfnamefont
  {X.~F.}\ \bibnamefont {Yang}},\ and\ \bibinfo {author} {\bibfnamefont
  {D.~T.}\ \bibnamefont {Yordanov}},\ }\bibfield  {title} {\bibinfo {title}
  {Charge radii measurements of exotic tin isotopes in the proximity of $n=50$
  and $n=82$},\ }\href {https://doi.org/10.1103/wbdx-k3cd} {\bibfield
  {journal} {\bibinfo  {journal} {Phys. Rev. Lett.}\ }\textbf {\bibinfo
  {volume} {135}},\ \bibinfo {pages} {222501} (\bibinfo {year}
  {2025})}\BibitemShut {NoStop}%
\bibitem [{\citenamefont {Mott}(1929)}]{mott29}%
  \BibitemOpen
  \bibfield  {author} {\bibinfo {author} {\bibfnamefont {N.~F.}\ \bibnamefont
  {Mott}},\ }\bibfield  {title} {\bibinfo {title} {The scattering of fast
  electrons by atomic nuclei},\ }\href {http://doi.org/10.1098/rspa.1929.0127}
  {\bibfield  {journal} {\bibinfo  {journal} {Proc. R. Soc. Lond.}\ }\textbf
  {\bibinfo {volume} {A124}},\ \bibinfo {pages} {425–442} (\bibinfo {year}
  {1929})}\BibitemShut {NoStop}%
\bibitem [{\citenamefont {Mott}(1932)}]{mott32}%
  \BibitemOpen
  \bibfield  {author} {\bibinfo {author} {\bibfnamefont {N.~F.}\ \bibnamefont
  {Mott}},\ }\bibfield  {title} {\bibinfo {title} {The polarisation of
  electrons by double scattering},\ }\href
  {http://doi.org/10.1098/rspa.1932.0044} {\bibfield  {journal} {\bibinfo
  {journal} {Proc. R. Soc. Lond.}\ }\textbf {\bibinfo {volume} {A135}},\
  \bibinfo {pages} {429–458} (\bibinfo {year} {1932})}\BibitemShut {NoStop}%
\bibitem [{\citenamefont {McKinley}\ and\ \citenamefont
  {Feshbach}(1948)}]{McKinley1948PhysRev.74.1759}%
  \BibitemOpen
  \bibfield  {author} {\bibinfo {author} {\bibfnamefont {W.~A.}\ \bibnamefont
  {McKinley}}\ and\ \bibinfo {author} {\bibfnamefont {H.}~\bibnamefont
  {Feshbach}},\ }\bibfield  {title} {\bibinfo {title} {The coulomb scattering
  of relativistic electrons by nuclei},\ }\href
  {https://doi.org/10.1103/PhysRev.74.1759} {\bibfield  {journal} {\bibinfo
  {journal} {Phys. Rev.}\ }\textbf {\bibinfo {volume} {74}},\ \bibinfo {pages}
  {1759} (\bibinfo {year} {1948})}\BibitemShut {NoStop}%
\bibitem [{\citenamefont {Doggett}\ and\ \citenamefont
  {Spencer}(1956)}]{Doggett1956PhysRev.103.1597}%
  \BibitemOpen
  \bibfield  {author} {\bibinfo {author} {\bibfnamefont {J.~A.}\ \bibnamefont
  {Doggett}}\ and\ \bibinfo {author} {\bibfnamefont {L.~V.}\ \bibnamefont
  {Spencer}},\ }\bibfield  {title} {\bibinfo {title} {Elastic scattering of
  electrons and positrons by point nuclei},\ }\href
  {https://doi.org/10.1103/PhysRev.103.1597} {\bibfield  {journal} {\bibinfo
  {journal} {Phys. Rev.}\ }\textbf {\bibinfo {volume} {103}},\ \bibinfo {pages}
  {1597} (\bibinfo {year} {1956})}\BibitemShut {NoStop}%
\bibitem [{\citenamefont {Johnson}\ \emph {et~al.}(1961)\citenamefont
  {Johnson}, \citenamefont {Weber},\ and\ \citenamefont
  {Mullin}}]{johnson1961}%
  \BibitemOpen
  \bibfield  {author} {\bibinfo {author} {\bibfnamefont {W.~R.}\ \bibnamefont
  {Johnson}}, \bibinfo {author} {\bibfnamefont {T.~A.}\ \bibnamefont {Weber}},\
  and\ \bibinfo {author} {\bibfnamefont {C.~J.}\ \bibnamefont {Mullin}},\
  }\bibfield  {title} {\bibinfo {title} {Coulomb scattering of polarized
  electrons},\ }\href {https://doi.org/10.1103/PhysRev.121.933} {\bibfield
  {journal} {\bibinfo  {journal} {Phys. Rev.}\ }\textbf {\bibinfo {volume}
  {121}},\ \bibinfo {pages} {933} (\bibinfo {year} {1961})}\BibitemShut
  {NoStop}%
\bibitem [{\citenamefont {Tolhoek}(1956)}]{Tolhoek1956}%
  \BibitemOpen
  \bibfield  {author} {\bibinfo {author} {\bibfnamefont {H.~A.}\ \bibnamefont
  {Tolhoek}},\ }\bibfield  {title} {\bibinfo {title} {Electron polarization,
  theory and experiment},\ }\href {https://doi.org/10.1103/RevModPhys.28.277}
  {\bibfield  {journal} {\bibinfo  {journal} {Rev. Mod. Phys.}\ }\textbf
  {\bibinfo {volume} {28}},\ \bibinfo {pages} {277} (\bibinfo {year}
  {1956})}\BibitemShut {NoStop}%
\bibitem [{\citenamefont {Sherman}(1956)}]{sherman56}%
  \BibitemOpen
  \bibfield  {author} {\bibinfo {author} {\bibfnamefont {N.}~\bibnamefont
  {Sherman}},\ }\bibfield  {title} {\bibinfo {title} {Coulomb scattering of
  relativistic electrons by point nuclei},\ }\href
  {https://doi.org/10.1103/PhysRev.103.1601} {\bibfield  {journal} {\bibinfo
  {journal} {Phys. Rev.}\ }\textbf {\bibinfo {volume} {103}},\ \bibinfo {pages}
  {1601} (\bibinfo {year} {1956})}\BibitemShut {NoStop}%
\bibitem [{\citenamefont {Uehling}(1935)}]{uehling35PhysRev.48.55}%
  \BibitemOpen
  \bibfield  {author} {\bibinfo {author} {\bibfnamefont {E.~A.}\ \bibnamefont
  {Uehling}},\ }\bibfield  {title} {\bibinfo {title} {Polarization effects in
  the positron theory},\ }\href {https://doi.org/10.1103/PhysRev.48.55}
  {\bibfield  {journal} {\bibinfo  {journal} {Phys. Rev.}\ }\textbf {\bibinfo
  {volume} {48}},\ \bibinfo {pages} {55} (\bibinfo {year} {1935})}\BibitemShut
  {NoStop}%
\bibitem [{\citenamefont {Fullerton}\ and\ \citenamefont
  {Rinker}(1976)}]{Fullerton1976PhysRevA.13.1283}%
  \BibitemOpen
  \bibfield  {author} {\bibinfo {author} {\bibfnamefont {L.~W.}\ \bibnamefont
  {Fullerton}}\ and\ \bibinfo {author} {\bibfnamefont {G.~A.}\ \bibnamefont
  {Rinker}},\ }\bibfield  {title} {\bibinfo {title} {Accurate and efficient
  methods for the evaluation of vacuum-polarization potentials of order
  {$Z\ensuremath{\alpha}$} and {$Z{\ensuremath{\alpha}}^{2}$}},\ }\href
  {https://doi.org/10.1103/PhysRevA.13.1283} {\bibfield  {journal} {\bibinfo
  {journal} {Phys. Rev. A}\ }\textbf {\bibinfo {volume} {13}},\ \bibinfo
  {pages} {1283} (\bibinfo {year} {1976})}\BibitemShut {NoStop}%
\bibitem [{\citenamefont {Mohr}\ \emph {et~al.}(1998)\citenamefont {Mohr},
  \citenamefont {Plunien},\ and\ \citenamefont {Soff}}]{mohr1998pr293-227}%
  \BibitemOpen
  \bibfield  {author} {\bibinfo {author} {\bibfnamefont {P.~J.}\ \bibnamefont
  {Mohr}}, \bibinfo {author} {\bibfnamefont {G.}~\bibnamefont {Plunien}},\ and\
  \bibinfo {author} {\bibfnamefont {G.}~\bibnamefont {Soff}},\ }\bibfield
  {title} {\bibinfo {title} {{QED} corrections in heavy atoms},\ }\href
  {https://doi.org/https://doi.org/10.1016/S0370-1573(97)00046-X} {\bibfield
  {journal} {\bibinfo  {journal} {Physics Reports}\ }\textbf {\bibinfo {volume}
  {293}},\ \bibinfo {pages} {227} (\bibinfo {year} {1998})}\BibitemShut
  {NoStop}%
\bibitem [{\citenamefont {Furry}(1951)}]{furry51}%
  \BibitemOpen
  \bibfield  {author} {\bibinfo {author} {\bibfnamefont {W.~H.}\ \bibnamefont
  {Furry}},\ }\bibfield  {title} {\bibinfo {title} {On bound states and
  scattering in positron theory},\ }\href
  {https://doi.org/10.1103/PhysRev.81.115} {\bibfield  {journal} {\bibinfo
  {journal} {Phys. Rev.}\ }\textbf {\bibinfo {volume} {81}},\ \bibinfo {pages}
  {115} (\bibinfo {year} {1951})}\BibitemShut {NoStop}%
\bibitem [{\citenamefont {Akhiezer}\ and\ \citenamefont
  {Berestetskii}(1965)}]{akhiezer65b}%
  \BibitemOpen
  \bibfield  {author} {\bibinfo {author} {\bibfnamefont {A.~I.}\ \bibnamefont
  {Akhiezer}}\ and\ \bibinfo {author} {\bibfnamefont {V.~B.}\ \bibnamefont
  {Berestetskii}},\ }\href@noop {} {\emph {\bibinfo {title} {Quantum
  Electrodynamics}}}\ (\bibinfo  {publisher} {Wiley Interscience},\ \bibinfo
  {address} {New York},\ \bibinfo {year} {1965})\BibitemShut {NoStop}%
\bibitem [{\citenamefont {Varshalovich}\ \emph {et~al.}(1988)\citenamefont
  {Varshalovich}, \citenamefont {Moskalev},\ and\ \citenamefont
  {Khersonskii}}]{Varshalovich1988QuantumTO}%
  \BibitemOpen
  \bibfield  {author} {\bibinfo {author} {\bibfnamefont {D.~A.}\ \bibnamefont
  {Varshalovich}}, \bibinfo {author} {\bibfnamefont {A.~N.}\ \bibnamefont
  {Moskalev}},\ and\ \bibinfo {author} {\bibfnamefont {V.~K.}\ \bibnamefont
  {Khersonskii}},\ }\href {https://doi.org/10.1142/0270} {\emph {\bibinfo
  {title} {Quantum Theory of Angular Momentum}}}\ (\bibinfo  {publisher} {World
  Scientific Publishing Co. Pte. Ltd.},\ \bibinfo {address} {Singapore},\
  \bibinfo {year} {1988})\BibitemShut {NoStop}%
\bibitem [{\citenamefont {Salvat}\ \emph {et~al.}(1995)\citenamefont {Salvat},
  \citenamefont {Fernández-Varea},\ and\ \citenamefont
  {Williamson}}]{salvat95}%
  \BibitemOpen
  \bibfield  {author} {\bibinfo {author} {\bibfnamefont {F.}~\bibnamefont
  {Salvat}}, \bibinfo {author} {\bibfnamefont {J.}~\bibnamefont
  {Fernández-Varea}},\ and\ \bibinfo {author} {\bibfnamefont {W.}~\bibnamefont
  {Williamson}},\ }\bibfield  {title} {\bibinfo {title} {Accurate numerical
  solution of the radial \mbox{S}chrödinger and \mbox{D}irac wave equations},\
  }\href {https://doi.org/https://doi.org/10.1016/0010-4655(95)00039-I}
  {\bibfield  {journal} {\bibinfo  {journal} {Computer Physics Communications}\
  }\textbf {\bibinfo {volume} {90}},\ \bibinfo {pages} {151} (\bibinfo {year}
  {1995})}\BibitemShut {NoStop}%
\bibitem [{\citenamefont {Berestetskii}\ \emph {et~al.}(1982)\citenamefont
  {Berestetskii}, \citenamefont {Lifshits},\ and\ \citenamefont
  {Pitaevskii}}]{landau4}%
  \BibitemOpen
  \bibfield  {author} {\bibinfo {author} {\bibfnamefont {V.}~\bibnamefont
  {Berestetskii}}, \bibinfo {author} {\bibfnamefont {E.}~\bibnamefont
  {Lifshits}},\ and\ \bibinfo {author} {\bibfnamefont {L.}~\bibnamefont
  {Pitaevskii}},\ }\href@noop {} {\emph {\bibinfo {title} {Quantum
  Electrodynamics}}},\ Course of theoretical physics\ (\bibinfo  {publisher}
  {Pergamon Press},\ \bibinfo {year} {1982})\BibitemShut {NoStop}%
\bibitem [{\citenamefont {Burke}(2011)}]{burke2011b}%
  \BibitemOpen
  \bibfield  {author} {\bibinfo {author} {\bibfnamefont {P.}~\bibnamefont
  {Burke}},\ }\href {https://doi.org/10.1007/978-3-642-15931-2} {\emph
  {\bibinfo {title} {R-matrix theory of atomic collisions}}}\ (\bibinfo
  {publisher} {Springer-Verlag},\ \bibinfo {address} {Berlin Heidelberg},\
  \bibinfo {year} {2011})\BibitemShut {NoStop}%
\bibitem [{\citenamefont {Lyashchenko}\ \emph {et~al.}(2020)\citenamefont
  {Lyashchenko}, \citenamefont {Vasileva}, \citenamefont {Andreev},\ and\
  \citenamefont {Voitkiv}}]{res2020}%
  \BibitemOpen
  \bibfield  {author} {\bibinfo {author} {\bibfnamefont {K.~N.}\ \bibnamefont
  {Lyashchenko}}, \bibinfo {author} {\bibfnamefont {D.~M.}\ \bibnamefont
  {Vasileva}}, \bibinfo {author} {\bibfnamefont {O.~Y.}\ \bibnamefont
  {Andreev}},\ and\ \bibinfo {author} {\bibfnamefont {A.~B.}\ \bibnamefont
  {Voitkiv}},\ }\bibfield  {title} {\bibinfo {title} {Qed theory of elastic
  electron scattering on hydrogen-like ions involving formation and decay of
  autoionizing states},\ }\href
  {https://doi.org/10.1103/PhysRevResearch.2.013087} {\bibfield  {journal}
  {\bibinfo  {journal} {Phys. Rev. Research}\ }\textbf {\bibinfo {volume}
  {2}},\ \bibinfo {pages} {013087} (\bibinfo {year} {2020})}\BibitemShut
  {NoStop}%
\bibitem [{\citenamefont {Angeli}\ and\ \citenamefont
  {Marinova}(2013)}]{ANGELI201369}%
  \BibitemOpen
  \bibfield  {author} {\bibinfo {author} {\bibfnamefont {I.}~\bibnamefont
  {Angeli}}\ and\ \bibinfo {author} {\bibfnamefont {K.}~\bibnamefont
  {Marinova}},\ }\bibfield  {title} {\bibinfo {title} {Table of experimental
  nuclear ground state charge radii: An update},\ }\href
  {https://doi.org/https://doi.org/10.1016/j.adt.2011.12.006} {\bibfield
  {journal} {\bibinfo  {journal} {Atomic Data and Nuclear Data Tables}\
  }\textbf {\bibinfo {volume} {99}},\ \bibinfo {pages} {69} (\bibinfo {year}
  {2013})}\BibitemShut {NoStop}%
\bibitem [{\citenamefont {Landau}\ and\ \citenamefont
  {Lifshitz}(1975)}]{landau1975V2}%
  \BibitemOpen
  \bibfield  {author} {\bibinfo {author} {\bibfnamefont {L.~D.}\ \bibnamefont
  {Landau}}\ and\ \bibinfo {author} {\bibfnamefont {E.~M.}\ \bibnamefont
  {Lifshitz}},\ }\href@noop {} {\emph {\bibinfo {title} {The Classical Theory
  of Fields}}},\ \bibinfo {edition} {4th}\ ed.,\ Vol.~\bibinfo {volume} {2}\
  (\bibinfo  {publisher} {Butterworth-Heinemann},\ \bibinfo {year}
  {1975})\BibitemShut {NoStop}%
\bibitem [{\citenamefont {Blundell}\ \emph {et~al.}(2021)\citenamefont
  {Blundell}, \citenamefont {De~Renzi}, \citenamefont {Lancaster},\ and\
  \citenamefont {Pratt}}]{Blundell2021muon}%
  \BibitemOpen
  \bibinfo {editor} {\bibfnamefont {S.~J.}\ \bibnamefont {Blundell}}, \bibinfo
  {editor} {\bibfnamefont {R.}~\bibnamefont {De~Renzi}}, \bibinfo {editor}
  {\bibfnamefont {T.}~\bibnamefont {Lancaster}},\ and\ \bibinfo {editor}
  {\bibfnamefont {F.~L.}\ \bibnamefont {Pratt}},\ eds.,\ \href
  {https://doi.org/10.1093/oso/9780198858959.001.0001} {\emph {\bibinfo {title}
  {Muon Spectroscopy: An Introduction}}}\ (\bibinfo  {publisher} {Oxford
  University Press},\ \bibinfo {year} {2021})\BibitemShut {NoStop}%
\bibitem [{\citenamefont {Burke}\ and\ \citenamefont
  {Mitchell}(1974)}]{Burke1974}%
  \BibitemOpen
  \bibfield  {author} {\bibinfo {author} {\bibfnamefont {P.~G.}\ \bibnamefont
  {Burke}}\ and\ \bibinfo {author} {\bibfnamefont {J.~F.~B.}\ \bibnamefont
  {Mitchell}},\ }\bibfield  {title} {\bibinfo {title} {Spin-polarization in the
  elastic scattering of electrons by one-electron atoms},\ }\href
  {https://doi.org/10.1088/0022-3700/7/2/007} {\bibfield  {journal} {\bibinfo
  {journal} {Journal of Physics B: Atomic and Molecular Physics}\ }\textbf
  {\bibinfo {volume} {7}},\ \bibinfo {pages} {214} (\bibinfo {year}
  {1974})}\BibitemShut {NoStop}%
\bibitem [{\citenamefont {Furry}(1934)}]{Furry1934PhysRev.46.391}%
  \BibitemOpen
  \bibfield  {author} {\bibinfo {author} {\bibfnamefont {W.~H.}\ \bibnamefont
  {Furry}},\ }\bibfield  {title} {\bibinfo {title} {Approximate wave functions
  for high energy electrons in coulomb fields},\ }\href
  {https://doi.org/10.1103/PhysRev.46.391} {\bibfield  {journal} {\bibinfo
  {journal} {Phys. Rev.}\ }\textbf {\bibinfo {volume} {46}},\ \bibinfo {pages}
  {391} (\bibinfo {year} {1934})}\BibitemShut {NoStop}%
\bibitem [{\citenamefont {Sommerfeld}\ and\ \citenamefont
  {Maue}(1935)}]{Sommerfeld1935}%
  \BibitemOpen
  \bibfield  {author} {\bibinfo {author} {\bibfnamefont {A.}~\bibnamefont
  {Sommerfeld}}\ and\ \bibinfo {author} {\bibfnamefont {A.~W.}\ \bibnamefont
  {Maue}},\ }\bibfield  {title} {\bibinfo {title} {Verfahren zur
  näherungsweisen anpassung einer lösung der schrödinger- an die
  diracgleichung},\ }\href {https://doi.org/10.1002/andp.19354140703}
  {\bibfield  {journal} {\bibinfo  {journal} {Ann. Phys.}\ }\textbf {\bibinfo
  {volume} {22}},\ \bibinfo {pages} {629} (\bibinfo {year} {1935})}\BibitemShut
  {NoStop}%
\bibitem [{\citenamefont {Bethe}\ and\ \citenamefont
  {Maximon}(1954)}]{Bethe1954PhysRev.93.768}%
  \BibitemOpen
  \bibfield  {author} {\bibinfo {author} {\bibfnamefont {H.~A.}\ \bibnamefont
  {Bethe}}\ and\ \bibinfo {author} {\bibfnamefont {L.~C.}\ \bibnamefont
  {Maximon}},\ }\bibfield  {title} {\bibinfo {title} {Theory of bremsstrahlung
  and pair production. i. differential cross section},\ }\href
  {https://doi.org/10.1103/PhysRev.93.768} {\bibfield  {journal} {\bibinfo
  {journal} {Phys. Rev.}\ }\textbf {\bibinfo {volume} {93}},\ \bibinfo {pages}
  {768} (\bibinfo {year} {1954})}\BibitemShut {NoStop}%
\bibitem [{\citenamefont {Elwert}\ and\ \citenamefont
  {Haug}(1969)}]{Elwert1969PhysRev.183.90}%
  \BibitemOpen
  \bibfield  {author} {\bibinfo {author} {\bibfnamefont {G.}~\bibnamefont
  {Elwert}}\ and\ \bibinfo {author} {\bibfnamefont {E.}~\bibnamefont {Haug}},\
  }\bibfield  {title} {\bibinfo {title} {Calculation of bremsstrahlung cross
  sections with sommerfeld-maue eigenfunctions},\ }\href
  {https://doi.org/10.1103/PhysRev.183.90} {\bibfield  {journal} {\bibinfo
  {journal} {Phys. Rev.}\ }\textbf {\bibinfo {volume} {183}},\ \bibinfo {pages}
  {90} (\bibinfo {year} {1969})}\BibitemShut {NoStop}%
\bibitem [{\citenamefont {Jakubassa-Amundsen}\ and\ \citenamefont
  {Yerokhin}(2013)}]{Jakubassa2013}%
  \BibitemOpen
  \bibfield  {author} {\bibinfo {author} {\bibfnamefont {D.~H.}\ \bibnamefont
  {Jakubassa-Amundsen}}\ and\ \bibinfo {author} {\bibfnamefont {V.~A.}\
  \bibnamefont {Yerokhin}},\ }\bibfield  {title} {\bibinfo {title}
  {Relativistic theory for radiative ionization of light atoms by heavy ions},\
  }\href {https://doi.org/10.1140/epjd/e2012-30483-7} {\bibfield  {journal}
  {\bibinfo  {journal} {The European Physical Journal D}\ }\textbf {\bibinfo
  {volume} {67}},\ \bibinfo {pages} {4} (\bibinfo {year} {2013})}\BibitemShut
  {NoStop}%
\bibitem [{\citenamefont {Subramanian}\ and\ \citenamefont
  {Devanathan}(1974)}]{Subramanian_1974}%
  \BibitemOpen
  \bibfield  {author} {\bibinfo {author} {\bibfnamefont {P.~R.}\ \bibnamefont
  {Subramanian}}\ and\ \bibinfo {author} {\bibfnamefont {V.}~\bibnamefont
  {Devanathan}},\ }\bibfield  {title} {\bibinfo {title} {Trace techniques for
  angular momentum operators},\ }\href
  {https://doi.org/10.1088/0305-4470/7/16/004} {\bibfield  {journal} {\bibinfo
  {journal} {Journal of Physics A: Mathematical, Nuclear and General}\ }\textbf
  {\bibinfo {volume} {7}},\ \bibinfo {pages} {1995} (\bibinfo {year}
  {1974})}\BibitemShut {NoStop}%
\end{thebibliography}
\end{document}